\pdfoutput=1 
\documentclass[fleqn,usenatbib]{mnras}
\usepackage{tabularx}
\usepackage{lineno}
\usepackage{graphicx}
\usepackage[para]{threeparttable}
\usepackage{ulem}
\usepackage{newtxtext,newtxmath}
\usepackage{array}
\usepackage[dvipsnames]{xcolor}

\usepackage{newtxtext,newtxmath}
\usepackage{array}
\newcolumntype{P}[1]{>{\centering\arraybackslash}p{#1}}

\usepackage[T1]{fontenc}
\usepackage{ae,aecompl}

\usepackage{graphicx}	
\usepackage{amsmath}	
\usepackage{float}

\title[Supernova-Cloud Interactions]{Shocking interactions of supernova remnants with atomic and molecular clouds - the interplay between shocks, thermal instability and gravity in the large cloud regime}

\author[M. M. Kupilas et al.]{
M. M. Kupilas,$^{1}$\thanks{py12mk@leeds.ac.uk}
J. M. Pittard,$^{1}$
C. J. Wareing,$^{1}$
S. A. E. G. Falle$^{2}$
\\
$^{1}$ School of Physics and Astronomy, University of Leeds, LS2 9JT, UK \\
$^{2}$ Department of Applied Mathematics, University of Leeds, LS2 9JT, UK \\
}

\date{Accepted 2022 April 04. Received 2022 March 17; in original form 2022 January 12}

\pubyear{2019}

\begin{document}
\setcounter{secnumdepth}{5}
\setcounter{tocdepth}{5}

\label{firstpage}
\pagerange{\pageref{firstpage}--\pageref{lastpage}}
\maketitle

\begin{abstract}

\noindent Using the adaptive mesh refinement code \textsc{mg}, we perform 3D hydrodynamic simulations of a supernova-cloud interaction in the "large cloud regime". The cloud is initially atomic and evolving due to the thermal instability (TI) and gravity. We study interactions in a "pre-TI" and "post-TI" stage when cold and dense clumps are present, and compare these results to idealised shock-cloud scenarios in the "small cloud regime", and a scenario without shocks. On aggregate, the supernova disruption is significantly weaker than that from an idealised shock due to the supernova impact being instantaneous, and not continuous. In both supernova-cloud interactions, we observe two shocks impact the cloud, followed by the development of a weak 10\,km\,s$^{-1}$ upstream flow on the cloud interface, and a global ambient pressure drop. When the cloud is still atomic, it expands due to this drop. Additionally, the TI is triggered at the front of the cloud, causing the formation of a cap-like structure with clumps embedded inside. The upstream flow converges in this region, resulting in a lobe-like cloud morphology. When the cloud is molecular, the transmitted shock disrupts the inter-clump material and causes the clumps' outer envelopes to expand slightly and form tail-like morphologies. These effects are less pronounced than those in our shock-cloud scenarios, and more pronounced that those in our un-shocked scenario. After $\sim$\,3.5\,Myrs, the effects from the supernova decay and the cloud returns to an almost indistinguishable state from an un-shocked cloud, in spite of the global ambient pressure drop. In neither supernova-cloud scenario do we see any local gravitational collapse.

\end{abstract}


\section{Introduction}

The interstellar medium (ISM) is composed of multiple phases, ranging from a hot ionized plasma (\textit{T}\,$\gtrsim$\,10$^6$\,K, \textit{n}\,$\lesssim$\,0.01\,cm$^{-3}$) to a cold molecular gas (\textit{T}\,$\sim$\,10\,--\,20\,K, \textit{n}\,$\gtrsim$\,100\,cm$^{-3}$) \citep{mckee1977theory,wolfire1995neutral,wolfire2003neutral}.
Stars form in molecular clouds, and thus to convert ISM material into a star involves a web of processes occurring on time and length scales spanning many orders of magnitude, ranging from galactic scales down to the scale of the individual star.
For a self-consistent theory of star formation, details of the physics in the different phases across the time and length scales must be connected and understood.
This is far from being the case, and many outstanding problems remain \citep[e.g. see][]{krumholz2014big}.
%
%

%
%
%
%
%
In the context of molecular cloud formation, the interplay between the physics of the thermal instability \citep[TI,][]{parker1953instability, field1965thermal}, gravity and magnetic fields was studied in detail by \citet[][hereafter WPFVL16]{wareing2016} in order to help elucidate the relative importance of these processes at different stages.
They performed idealised 2D and 3D simulations of an initially quiescent
cloud seeded with $\pm$10 per cent density perturbations around an equilibrium state of \textit{n}$_\textup{cl}$\,=\,1.1\,cm$^{-3}$, in the warm, neutral atomic unstable phase.
They found that the TI alone was able to create clumps, and filaments in the magnetic field case, whose properties connected well with observations.
The models were designed to incrementally study the relative impact of different physics, and continued to increase in complexity including feedback \citep{wareing2016magnetohydrodynamic,wareing2017hydrodynamic,wareing2018new}, and larger cloud masses with higher resolution \citep[][hereafter WFP19]{wareing2019} and magnetic fields \citep{wareing2021striations}.

In \citet[][Paper I]{kupilas10.1093/mnras/staa3889} we looked at an extension of the 3D hydrodynamical scenarios in WPFVL16 by including a shock. 
The simulations were set up as a plane-parallel shock-cloud system commonly seen in the literature \citep[e.g.][]{stone1992three,klein1994hydrodynamic,nakamura2006hydrodynamic,van2007shock,pittard2009turbulent,Pittard2011Tails,kinoshita2021star,kanjilal2021growth} and the relative importance of the TI, gravity and the shock impact was explored. 
Two scenarios were studied consisting of a shock-cloud interaction with a "pre-TI" warm atomic cloud, and a "post-TI" molecular cloud consisting of cold and dense (\textit{T}\,$\sim$\,50\,--\,160\,K, \textit{n}\,$\sim\,$100\,--\,1000\,cm$^{-3}$) clumps embedded in a warm and diffuse (\textit{T}\,$\sim$\,1000\,K, \textit{n}\,$\approx$\,0.8\,cm$^{-3}$) interclump gas. 

In both cases, the constant impact due to the post-shock flow compressed the cloud significantly and accelerated the global gravitational collapse of the cloud. 
Evidence of local gravitational collapse was seen also, occurring on a $\sim$\,5\,Myr timescale in both scenarios.

Planar shocks like that in Paper I and the literature are usually approximations of more complex flow patterns such as those resulting from supernova remnants, a so-called "small-cloud approximation" \citep[e.g. see][]{klein1994hydrodynamic}. It is prudent therefore to investigate deviations from such an approximation, and explore more realistic scenarios in a "large cloud regime", where the idealised shock is replaced by a supernova explosion.
Indeed, there exists a history of studies with a vast range of initial conditions that consider the impact of supernova explosions on molecular clouds and the wider ISM \citep[e.g.][]{wada2000formation,joung2006turbulent, rogers2013feedback,rogers2014feedback,padoan2016supernova,walch2015energy,li2015supernova, seifried2018molecular,lu2020effect}.
To our knowledge however, a systematic comparison between the physical behaviour of an idealised shock-cloud and supernova-cloud set-up has not yet been performed.
This is the goal of the current paper, and it is structured as follows.
In Section \ref{sec:Methods} we present details of the model, introduce the specific scenarios, define the small, medium and large cloud regimes, and outline the computational details. In Section \ref{sec:results} we present our results, followed by a brief discussion of physical implications in Section \ref{sec:discuss}, and we conclude our work in Section \ref{sec:conclucions}.

\section{Methods} \label{sec:Methods}
\subsection{Numerical method}

This work presents 3D hydrodynamical simulations of the interaction of a supernova remnant with a cloud that is evolving due to the thermal instability and gravity, and the interactions are compared to their analogous planar shock-cloud and un-shocked scenarios studied in Paper I.
All calculations were performed using the finite-volume adaptive mesh refinement (AMR) code \textsc{mg} \citep{Falle1991,falle2005amr,hubber2013convergence}. 
The code solves the Euler equations of hydrodynamics with free boundary conditions, is second order in space and time, employs a hybrid cooling function and treats gravity by solving the Poisson equation using the full-approximation multigrid technique.
For the reader interested in the details of the numerical scheme, they can be found in \citet{Falle1991}, WPFVL16 and Paper I.
We note however that in Paper I the simulations were initially ran with the Godunov (GOD) Riemann solver \citep{godunov1959difference} until the gradients in the domain were too high for the simulations to continue. At that point the solver was switched to Kurganov-Tadmor (KT) \citep{kurganov2000new}. In the current paper, the KT solver is chosen from the beginning.

\subsection{Model}

In this work we adopt the same initial model as scenario 3 in WPFVL16 and that in Paper I. Namely, 17\,000\,M$_\odot$ of diffuse material is initialised in a sphere with radius \textit{r}\,=\,50\,pc, which we refer to as the cloud (see figure 1 in Paper I).
The cloud lies at the origin of a Cartesian domain with a numerical extent of -3\,$<$\,\textit{xyz}\,$<$\,3. 
In our model a unit length corresponds to 50\,pc, resulting in a domain with 300\,pc extent on each side.
8 levels of AMR are used, with the grid fixed to maximum resolution within a sphere with radius \textit{r}\,=\,75\,pc centered at the origin to accurately capture the behaviour of the thermal instability (TI).
The grid refines and de-refines dynamically outside this region, with level 5 chosen as the minimum base level.
For the supernova scenarios, the region outside \textit{r}\,=\,75pc is then forced to be resolved down to the 5th level to manage the computational cost associated with the grid creating levels and cells as the remnant expands into regions that are of no interest to the interaction. 
Note that for a fully refined grid, the maximum number of grid cells would be 1024$^3$, however due to the AMR, $\lesssim$ 10 per cent of this is used. The maximum resolution on the finest grid is $\Delta$\textit{x}\,=\,0.29\,pc. A discussion of the dependence of the thermal instability on the grid resolution and our arguments as to why it is acceptably resolved in our simulations are presented in the appendix of Paper I.
The cloud is seeded with $\pm$\,10 per cent random density variations around \textit{n}$_\textup{cl}$\,=\,1.1\,cm$^{-3}$ in the thermally unstable phase.
The pressure for this is then \textit{P/k}\,=\,4800\,$\pm$\,300\,K\,cm$^{-3}$.
We note that we assume a mean particle mass of 2.0\,$\times$\,10$^{-24}$\,g.
Just like in Paper~I, all material in the domain is initially quiescent and the cloud is embedded in a medium with a density contrast $\chi$\,=\,50 ($\chi$\,=\,\textit{n}$_\textup{cl}$/\textit{n}$_\textup{amb}$) setting the ambient medium \textit{n}$_\textup{amb}$\,=\,0.022\,cm$^{-3}$. 
An advected scalar $\alpha$ is set to 1 in the cloud and 0 everywhere else, and the energy source term is switched on only in regions where $\alpha$\,$>$\,0.9.
The value of the scalar and all the other fluid states transition monotonically between the cloud and ambient values over an interface region with thickness of $\sim$\,5 cells. 
At the moment of supernova/shock injection, the surroundings are remapped to have \textit{n}$_\textup{amb}$\,=\,0.0022 and \textit{T}$_\textup{amb}$\,=\,2.19\,$\times$\,10$^6$\,K ($\chi$\,=\,500) so that the energy source term can be switched on in the whole domain whilst the surroundings remain adiabatic due to the large cooling times of the order \textit{t}$_\textup{cool}$\,$\gtrsim$\,100\,Myr at such temperatures.

\subsubsection{Scenarios}
Five scenarios are considered in this paper: \textit{NoShock}, \textit{S1}, \textit{S2}, \textit{SN1}, \textit{SN2}. The first three, \textit{NoShock}, \textit{S1} and \textit{S2} are effectively the cases presented in Paper I. 
For self-consistency, the data from Paper I was not used and instead the models were re-simulated using the KT solver for accurate comparisons to the new \textit{SN1} and \textit{SN2} scenarios.
Just like in Paper I, the \textit{NoShock} scenario follows the evolution of the initial cloud for a free-fall timescale of \textit{t}\,$\approx$\,35\,Myr without any additional disturbances.
\textit{S1} is then the \textit{12Shock} analog, where a planar shock is introduced at 11.78\,Myr into the evolution of \textit{NoShock}.
\textit{SN1} then corresponds directly to \textit{S1}, with the idealised planar shock being replaced by a supernova explosion introduced at the exact same time.
\textit{S2} is then the \textit{24Shock} analog, however this time a shock is introduced at 26.5\,Myr into the evolution of \textit{NoShock} instead of 24\,Myr.
The reason for this later injection time is due to the KT solver being more diffusive than the GOD solver, causing the thermal instability to take longer to trigger in the re-simulated \textit{NoShock} scenario.
This causes the phase transition to occur $\sim$\,2\,Myr later than in Paper I, and as such we choose a later time for shock/supernova injection in the \textit{S2/SN2} scenarios.
\begin{figure}
\includegraphics[width=0.5\textwidth]{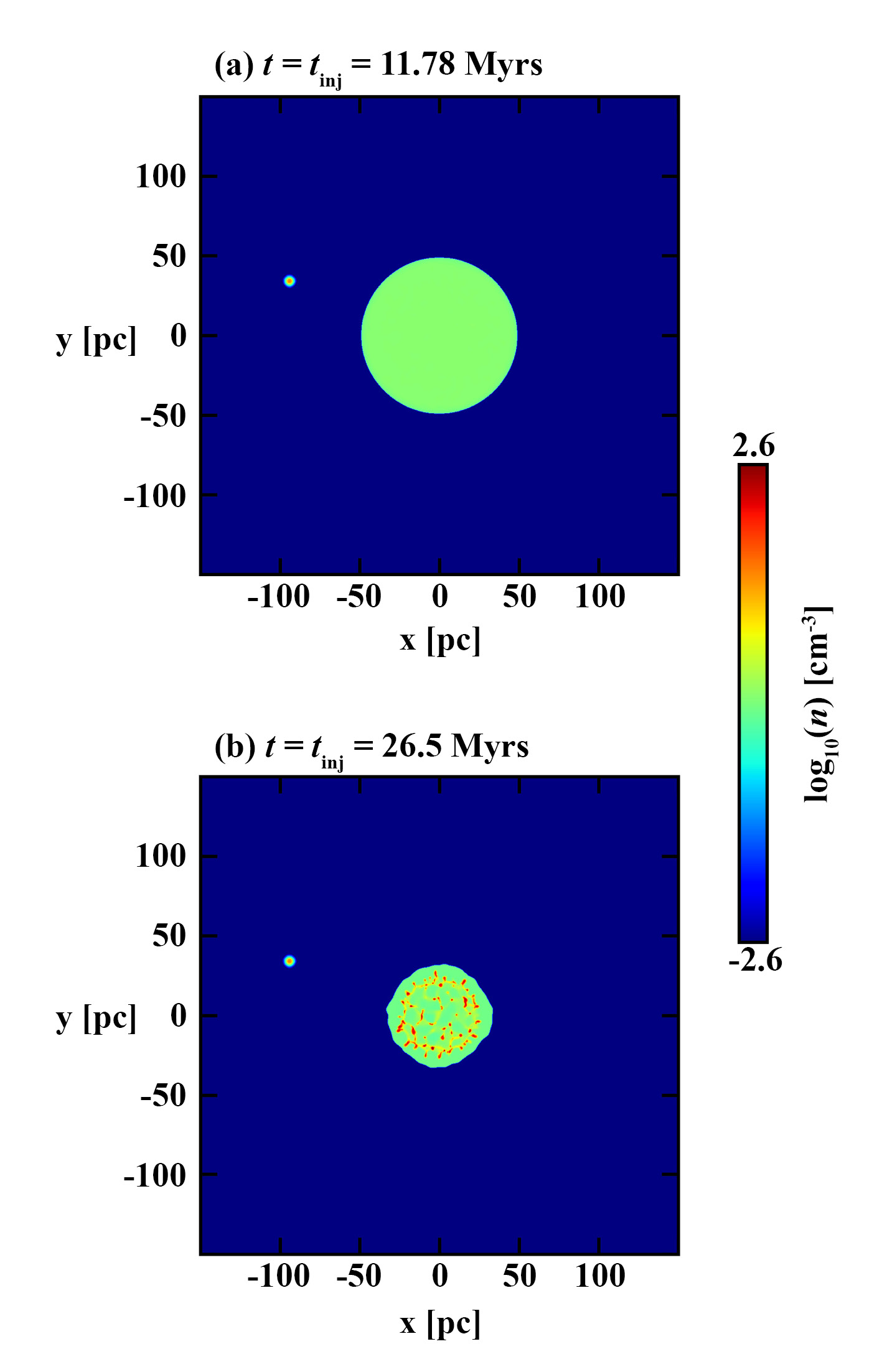}
\caption{Shown are snapshots of the \textit{z}\,=\,0 plane of the density logarithm of (a): The cloud at injection time \textit{t}$_\textup{inj}$\,=\,11.78\,Myr for \textit{SN1}/\textit{S1} scenarios and (b): Cloud at injection time \textit{t}$_\textup{inj}$\,=\,26.5\,Myr for \textit{SN2}/\textit{S2} scenarios. The location of the explosion can be seen in both snapshots.}
\label{fig:IC}
\end{figure} 
\begin{center}
\begin{table}
	\caption{Summary of all the scenarios studied in this work. The columns show the scenario name, the source of impact onto the cloud, the distance from the blast origin to the nearest cloud edge \textit{R}$_\textup{c}$, the Mach number (\textit{M}) of the disturbance shock immediately upon impact, and the injection time 
	(\textit{t}$_\textup{inj}$) of the disturbance. Scenarios \textit{NoShock}, \textit{S1} and 
	\textit{S2} are effectively \textit{NoShock}, \textit{12Shock} and \textit{24Shock} from 
	Paper I which have been re-simulated for this study for self-consistent comparisons. 
	Scenarios \textit{SN1} and 
	\textit{SN2} introduce supernova explosions at the same time as the shock in \textit{S1/S2}.}
    \label{table:models}
	\begin{tabularx}{\columnwidth}{XXXXX} 
		\hline
		Scenario & Disturbance & \textit{R}$_\textup{c}$ [pc] & $\textit{M}$ (impact) & \textit{t}$_\textup{inj}$ [Myr] \\ 
		\hline
		\textit{NoShock} & None         & \textit{N/A} & N/A          & N/A    \\
		\textit{S1}      & Shock &  $\infty$ & 1.5          & 11.78  \\
		\textit{SN1}     & Supernova    & 50 & $\approx$\,7* & 11.78  \\
		\textit{S2}      & Shock & $\infty$ & 1.5          & 26.5   \\
		\textit{SN2}     &Supernova    & 70 & $\approx$\,7* & 26.5   \\
		\hline

	\end{tabularx}
    \begin{tablenotes}
  
    \item [*] Due to a 10\,M$_\odot$ and 10$^{\tiny{51}}$\,erg explosion located at the polar coordinate (\textit{r}, $\theta$)\,=\,(100\,pc, 160$^\circ$) in the \textit{z}\,=\,0 plane. $\theta$ is measured from the positive \textit{x} axis.
    
    \end{tablenotes}
\end{table}
\end{center}
%
%
%

For our idealised shock-cloud models we follow an assumption known as the \textit{small cloud approximation} \citep[][hereafter KMC94]{klein1994hydrodynamic}, a condition that enables a physical set-up where an incoming shock onto a cloud has no curvature in its structure, and the shock-driving pressure is time independent. Implications of such an approximation can be understood in terms of the time-dependence of the shock-driving pressure, discussed in KMC94, and formally analysed in \citet[][hereafter MHST87]{mckee1987structure}.
In their theory of weakly time-dependent interstellar shocks, MHST87 introduce the pressure variation timescale \textit{t}$_\textup{p}$, and find that it approximates well as 
\begin{equation}
    t_\textup{p} \simeq
    0.1\frac{R_\textup{c}}{v_\textup{b}},
\end{equation}
\noindent where \textit{R}$_\textup{c}$ is the distance of the nearest part of the surface of the cloud from the blast epicentre, and 
\textit{v}$_\textup{b}$ = \textit{dR}$_\textup{b}$\textit{/dt} is the velocity of the blast wave.
Two other timescales of interest are the shock-crossing timescale

\begin{equation}
    t_\textup{sc} = \frac{2r_\textup{cl}}{v_\textup{b}},
\end{equation} 

\noindent i.e. the time taken for the external blast wave shock to sweep over the cloud, and the cloud-crushing timescale

\begin{equation}
    t_\textup{cc} = \frac{\chi^{1/2}r_\textup{cl}}{v_\textup{b}},
\end{equation}

\noindent i.e. the time taken for the transmitted cloud shock with velocity \textit{v}$_\textup{b,cl}$\,=\,\textit{v}$_\textup{b}$/$\chi^{1/2}$ to traverse the cloud radius \textit{r}$_\textup{cl}$ and crush the cloud.

Comparison of these timescales allows the definition of three regimes (see e.g. KMC94 and references therein).
The \textit{small cloud regime} requires the cloud to be sufficiently small (or blast to be sufficiently large) for \textit{t}$_\textup{p}$ to satisfy $t_\textup{cc} \ll t_\textup{p}$, thus by extension for \textit{r}$_\textup{cl}$ to satisfy
\begin{equation}
r_\mathrm{cl} \ll 0.1\frac{R_\mathrm{c}}{\chi^{1/2}}.
\end{equation}


\noindent The \textit{medium cloud regime} requires the cloud to be a size such that the blast wave does not change significantly as it sweeps over the cloud, but does change as the cloud is crushed. This requires \textit{t}$_\textup{p}$ to satisfy $t_\textup{cc} \gtrsim t_\textup{p} \gtrsim t_\textup{sc}$, and by extension for \textit{r}$_\textup{cl}$ to satisfy
\begin{equation}
0.1\frac{R_\textup{c}}{\chi^{1/2}} \lesssim r_\textup{cl}\,\lesssim\,0.05R_\textup{c}.
\end{equation}

\noindent Finally, the \textit{large cloud regime} requires that the cloud is large enough (or blast small enough) that the blast wave ages significantly as it sweeps over the cloud, resulting in vastly weaker compression at the rear of the cloud than the front, such that \textit{t}$_\textup{p}$ satisfies $t_\textup{sc}\,>\,t_\textup{p}$ and \textit{r}$_\textup{cl}$ satisfies
\begin{equation}
r_\textup{cl} > 0.05R_\textup{c}.
\end{equation}

For the \textit{S1}/\textit{S2} scenarios, an $\textit{M}$\,=\,1.5 shock is artificially imposed on the grid and boundary cells lying within the post-shock region are fixed to the post-shock values.
The shock front is located at the polar coordinate of (\textit{r},~$\theta$)\,=\,(50\,pc,~160$^\circ$) in the \textit{z}\,=\,0 plane, placing the shock immediately on the cloud's edge. The angle is measured counterclockwise from the positive \textit{x}-axis, and is chosen to minimise the effects of the Quirk instability, or the so-called \textit{carbuncle phenomenon}, that can cause perturbations along the shock symmetry axis, with strongest perturbation occurring when aligned with the grid \citep{quirk1997contribution,elling2009carbuncle}.
%
%
%

For our choice of parameters, \textit{S1/S2} scenarios can be considered to have 
R$_\textup{c}\,=\,\infty$, 
placing them in the small cloud regime as initially assummed.

For the \textit{SN1}/\textit{SN2} scenarios, a supernova is injected at the polar coordinate of (\textit{r}, $\theta$)\,=\,(100\,pc, 160$^\circ$) in the \textit{z}\,=\,0 plane on the grid located such that the first impact on the cloud is felt at the same location as the first shock impact in \textit{S1/S2}. 
The injection volume is artificially refined to the highest AMR level and subsequently chosen to fill a region with a radius of 5 cells, i.e. 1.5\,pc, into which 10\,M$_\odot$ of mass and 10$^{51}$\,erg of thermal energy are injected over a 500\,\textit{y}r period, roughly the time it would take for a remnant to expand to the injection volume. Note that immediately upon impact, the forward shock of the supernova remnant has $\textit{M}$\,$\approx$\,7.
The clouds at 11.78\,Myr and 26.5\,Myr are shown in figure \ref{fig:IC} where the region of supernova injection can also be seen. 

For \textit{SN1}, \textit{R}$_\textup{c}$\,$\approx$\,50\,pc which gives 0.05\textit{R}$_\textup{c}$\,=\,2.5\,pc. As in this scenario \textit{r}$_\textup{cl}$\,=\,50\,pc, it places this model firmly in the large cloud regime. For \textit{SN2}, \textit{R}$_\textup{c}$\,$\approx$\,70\,pc as the evolution results in the cloud shrinking slightly due to the formation of clumps and gravitational contraction, reducing \textit{r}$_\textup{cl}$ to a size of 30\,pc. Nevertheless, 0.05\textit{R}$_\textup{c}$ evaluates to 3.5\,pc, also placing this model firmly in the large cloud regime. Thus our new models showcase interactions in two contrasting, and very different regimes, with the large cloud regime showing a much more realistic interaction.
A summary of the model parameters is presented in Table \ref{table:models}.
\begin{figure*}
\includegraphics[width=\textwidth]{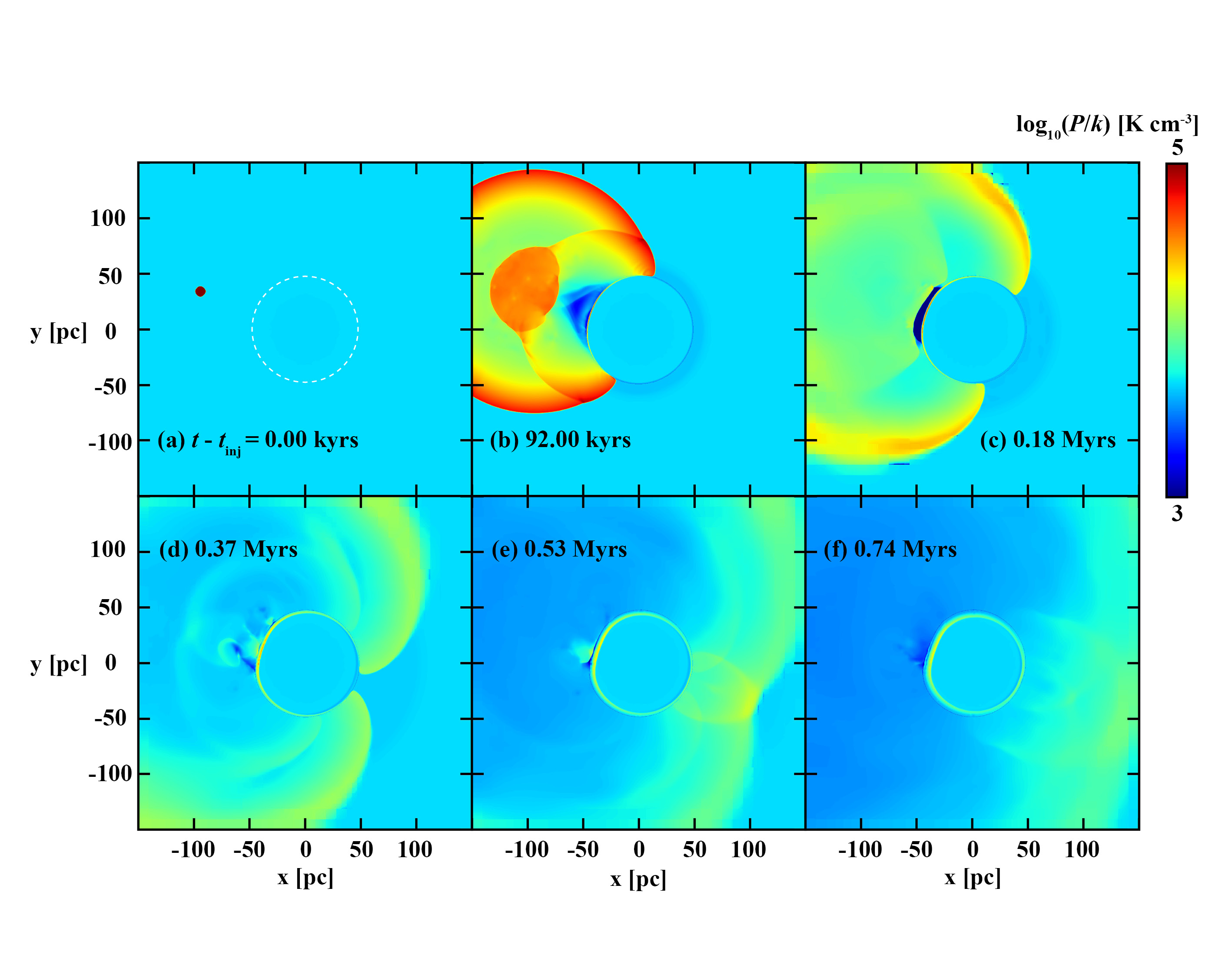}
\caption{Shown are pressure logarithm plots in the \textit{z}\,=\,0 plane of the \textit{SN1} scenario. The cloud edge cannot be seen initially in panel (a) and is thus marked by the white dashed line. 
The time in each panel corresponds to time since supernova injection \textit{t}\,-\,\textit{t}$_\textup{inj}$, with \textit{t}$_\textup{inj}$\,=\,11.78\,Myr. The final snapshot shows the moment immediately prior to the primary shock moving off the grid, and at the time the upstream flow around the cloud is being established. }
\label{fig:SN1pg}
\end{figure*} 

\section{Results}\label{sec:results}
%
        
        
        
        
        
%
%
%
%
%
%
%
%
%
%
%
In this section we present our results.
First, we present the \textit{SN1} scenario and discuss the evolution outside the cloud followed by a discussion of the internal evolution.
Secondly we present the \textit{SN2} scenario by again looking at the external and internal evolution separately.
%
%
%
\subsection{Interaction with atomic cloud - \textit{SN1}}

\noindent\newline
%
%
%
%
%
%
%
%
%
%
%
Throughout the discussion, comparisons are made to the planar shock scenario \textit{S1}, whose detailed evolution is presented in Paper I.
In addition, where appropriate, the \textit{NoShock} scenario is referenced.
The 5.16\,Myr timescale, where we choose to stop our analysis, corresponds to the time of the first \textit{S1} snapshot that showed evidence of local gravitational collapse.
%
%
%

\subsubsection{Dynamics outside the cloud}\label{sec:SN1external}
%
%
%
%
%
%
%
%
%
%
%
%
%

%
%

%
%
%
%
%
%
%
%
%
%
%
%
%

%
%
%
%
To illustrate the dynamics outside the cloud we first show in figure \ref{fig:SN1pg} the pressure profile of the domain in the \textit{z}\,=\,0 plane.
The first 0.74\,Myr are shown, as the most interesting features in the ambient medium are seen during this timescale. 
Figure \ref{fig:SN1pg}(a) shows a snapshot immediately post supernova injection.
A spherical region with a 50\,pc radius centered at the origin contains 17\,000\,M$_\odot$ of material, whose density is a factor $\chi$\,=\,500 greater than the surroundings with \textit{n}$_\textup{amb}$\,=\,0.0022\,cm$^{-3}$. 
%
%
A region with a radius \textit{r}\,=\,1.5\,pc contains 10\,M$_\odot$ and 10$^{51}$\,erg of thermal energy, and as such is strongly over-pressured with respect to its surroundings, with internal pressures exceeding 10$^{10}$\,K\,cm$^{-3}$, 7 orders of magnitude larger than the ambient pressure of approximately 4800\,K\,cm$^{-3}$.
A powerful explosion thus follows, accelerating stationary material to speeds exceeding 10$^4$\,km\,s$^{-1}$.
%

%
%
%
Before describing the interaction, we highlight a number of noticeable numerical effects. 
Firstly, the cloud in Fig.\,\ref{fig:SN1pg}(a) has a radius \textit{r}\,=\,50\,pc, however only a smaller spherical region with a radius \textit{r}\,$\approx$\,25\,pc can be seen.
This corresponds to where the internal pressure has dropped slightly due to the early action of the thermal instability. 
%
This behaviour is expected and acceptable, however we would expect to see the whole cloud behave this way.
Instead what we see is a numerical effect on the boundary resulting from how the source term was initially set up (i.e. switched off for regions where the scalar $\alpha$\,$<$\,0.9, with $\alpha$\,=\,1 inside the cloud and $\alpha$\,=\,0 outside).
Material below $\alpha$\,$<$\,0.9 does not cool, and as such a gradient develops starting at the interface (where $\alpha$\,$=$\,0.9) from the value of the ambient pressure (\textit{P/k}\,$\approx$\,4800\,K\,cm$^{-3}$) to the lowest value of the cloud centre (\textit{P/k}\,$\approx$\,4750\,K\,cm$^{-3}$), and accounts for why the full cloud is not visible in panel (a).
Where the edge actually is has been marked with a dashed white line.
%
%
Whilst this effect is noticeable, it accounts for a maximum of 1 per cent fractional difference in the fluid variables, and does not affect the dynamics.
Additionally, an outwardly propagating expansion wave can also be seen at the far side of the cloud, most clearly in panels (b) and (c).
This results from the mapping of the ambient medium from $\chi$\,=\,50 to $\chi$\,=\,500,
and amounts to a fractional difference in the fluid state of $\sim$\,5 per cent.

%
Again, this has little effect on the dynamics especially since these differences become overwhelmed by the supernova impact shortly thereafter.
The impact with the cloud occurs $\sim$\,40\,kyrs after explosion.
%
At this point, $\sim$\,33\,M$_\odot$ of material has been swept up, approximately 3\,$\times$\,the mass of the ejecta, and as such the profile immediately prior to impact is Sedov-Taylor-like.
%
The shock has a speed $v_\textup{s}$\,=\,1087\,km\,s$^{-1}$ upon impact, and given a pre-shock sound speed \textit{c}\,=\,160\,km\,s$^{-1}$, this amounts to an impact shock with Mach number \textit{M}\,=\,$v_\textup{s}/c$\,$\approx$\,7, roughly a factor $\sim$\,4.5\,$\times$ stronger than the \textit{M}\,=\,1.5 planar shock in \textit{S1}.
A reverse shock from the blast moves towards the epicentre with \textit{v}$_\textup{rs}$\,=\,450\,km\,s$^{-1}$, and a contact discontinuity separates the forward and reverse shocks.
%
Note that a purely analytical Sedov-Taylor profile does not have these two features - it is assummed that all waves have settled into a steady state such that a self-similar analytical solution is appropriate.

In our simulations the profile behind the forward shock matches well with the analytical Sedov-Taylor profile.
%
%
%


%
%
%
The profile interacts with the cloud as the remnant passes over it, transmitting a shock into the cloud and reflecting a shock back into the ambient medium. 
The reflected shock propagates towards the blast epicentre, and leaves behind a low pressure void due to a trailing expansion.
Meanwhile, 
the reverse shock from the initial blast reflects at the epicentre, and as it propagates back outwards, interacts with the cloud-reflected shock, generating an additional set of disturbances. 
This shock subsequently approaches the cloud and causes a secondary impact.
We refer to the cloud edge where this impact occurs as the front of the cloud, the first shock as the primary shock and the second shock as the secondary shock.
%
%
%


%
%
Fig.\,\ref{fig:SN1pg}(b) shows a snapshot at 92\,kyrs.
The primary shock has passed almost half of the cloud, and a clear distortion in its interior structure is seen in the region that interacts with the reflected shock. 
The primary shock diffracts around the cloud which further distorts its structure. 
A low pressure region ahead of the cloud can be seen, which forms due to the expansion of gas behind the reflected primary shock.
The gas at the front of the cloud continues to expand during the secondary shock approach, and as such there is a strong and sharp pressure contrast between this region and its surroundings - as evident in Fig.\,\ref{fig:SN1pg}(c) which shows the moment preceding the secondary impact.
By Fig.\,\ref{fig:SN1pg}(d) the secondary impact has occurred, resulting in turbulent eddies forming at the front of the cloud and a set of reflected waves propagating towards the left \textit{x} boundary.
The level of which this may seed turbulence is left to a future work.
Both shocks can now be seen sweeping over the cloud too, and the primary shock has nearly converged at the rear.
%
%
%


%
%
%
%
%
%
%
%
%
%
%
%
By Fig.\,\ref{fig:SN1pg}(e), the primary shock has converged and the secondary shock is trailing close behind.
%
As the primary shock converges, an interaction region is generated with portions of the resulting waves propagating both downstream and upstream, with the upstream waves interacting with the secondary shock as it sweeps over the cloud.
%

By Fig.\,\ref{fig:SN1pg}(f) both shocks have passed over the cloud, and are now at the domain edge. 
The secondary shock has now also converged at the rear, resulting in a secondary set of waves with both downstream and upstream components.
Around the edges of the cloud the upstream components generate a flow which then itself converges at the front, exaggerating the compression from the initial impact.
%
%
%


%
%
%
        
%
%
%
%
%
After Fig.\,\ref{fig:SN1pg}(f), the ambient medium evolves due to existing pressure gradients and the remaining flows on the grid. 
The ambient pressure in panel (f) can already be seen to be lower than most of the cloud's edge, and this continues to drop as the model evolves. 
The portions of the cloud that can respond to these changes, i.e. those that have been shocked, gradually expand in an attempt to balance the ambient pressures and the post-shock pressures in the cloud. 
This is halted however at the front of the cloud due to the ram pressure impact of the flow that converges there. 
Thus the outer envelope of the cloud expands everywhere but the front, resulting in a lobe-like morphology with the front being pinched inwards.
%
%
%


%
%
%
As early as the first 1 Myr of evolution, considerable differences between the \textit{SN1} and \textit{S1} scenario can be seen.
Two main ones can be identified however which exaggerate as the simulations evolve, and account for all of the differences seen between our shock-cloud and supernova-cloud scenarios.
Firstly, for an idealised, infinitely extending planar shock, the flow behind it is continuously replenished, resulting in disruption of the cloud throughout its entire evolution. 
Secondly, after shock passage, the ambient pressure is relatively time-independent, i.e after the medium is shocked and its pressures exceed that of the cloud, it stays that way, and provides continuous compression. 
%
These two features are present in all adiabatic, planar shock-cloud systems, and thus a more realistic model like ours can usefully be contrasted against them. 

\begin{figure*}
\includegraphics[width=\textwidth]{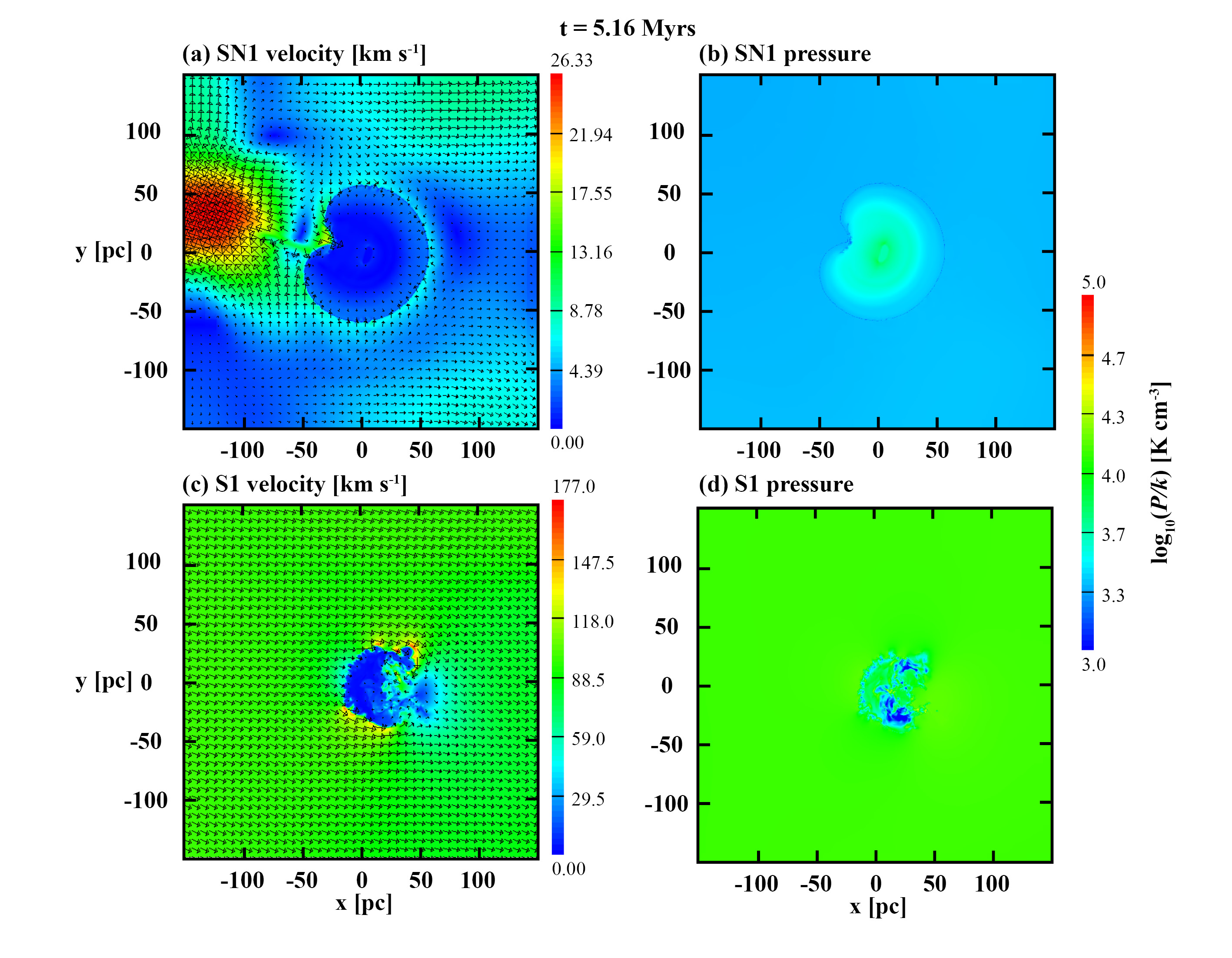}
\caption{Shown are slices in the \textit{z}\,=\,0 plane showing the velocity (panels a and c) and pressure (panels b and d) of \textit{SN1} (panels a and b) and \textit{S1} (panels c and d) at \textit{t}\,=\,5.16\,Myrs - the final snapshot presented in this work and the timescale at which the \textit{S1} scenario witnessed local gravitational collapse.}
\label{fig:SN1velpg}
\end{figure*} 


    
        
        
    

        
            
        
            
%
%

%
%
%
%
The impact of this in our models can clearly be seen in the last snapshot at 5.16 Myrs in figure \ref{fig:SN1velpg} which shows the pressure and velocity field in the \textit{z}\,=\,0 plane.
In Fig.\,\ref{fig:SN1velpg}b the ambient thermal pressure can be seen to have decreased down to $\sim$\,4000\,K\,cm$^{-3}$, approximately 80\,\% of its original value of 4800\,K\,cm$^{-3}$. 
The final \textit{S1} pressure in comparison is $\sim$\,10$^4$\,K\,cm$^{-3}$, a factor of 2.5 larger than the final \textit{SN1} pressure, and has not dropped since the initial shock passage.
The velocity difference is even larger, with impact amplitudes on the order of 10\,km\,s$^{-1}$ in \textit{SN1} (Fig.\,\ref{fig:SN1velpg}a) and 100\,km\,s$^{-1}$ in \textit{S1} (Fig.\,\ref{fig:SN1velpg}c), resulting in ram pressure impact that is $\sim$\,100 times greater ($\rho |\textbf{u}|^2|_{\textit{\tiny{S1}}}/\rho |\textbf{u}|^2|_{\textit{\tiny{SN1}}}$, $|\textbf{u}|^2$\,=\,$u_x^2 + u_y^2 + u_z^2$) in \textit{S1} than in \textit{SN1}.
Thus in aggregate, the supernova has a profoundly weaker compressive effect on the cloud, and is mainly responsible for creating external conditions that couple strongly to the cloud behaviour and result in cloud expansion.
%
%
%

%
%
%
We note that it is not entirely clear to what extent boundary effects impact the ambient evolution.
Many waves are expected due to back and forth reflections within the remnant \citep[e.g.][]{cioffi1988dynamics}.
As seen in Fig.\,\ref{fig:SN1pg}, the shocks move off the grid soon after 0.75\,Myrs and so these reflections cannot occur.
We would not however expect the fluid state to be significantly altered by their presence, and 1D spherically symmetric tests of supernova explosions with equivalent parameters as our models suggest this to be true.
In fact, in our 1D tests, once the reverse shock reflects at the blast epicentre, it is unable to catch up to the forward shock to reflect backwards once more.
An additional concern is the 26\,km\,s$^{-1}$ flows near the left \textit{x} boundary. 
Boundary effects are likely responsible for these flows as they do not align with the spherical symmetry of the initial blast.
However as they are not near the cloud and flowing off the grid, we do not expect them to affect the cloud.
Finally, the internal pressure of any supernova remnant drops as it expands.
As we see such a pressure drop occur in our models, we are confident that our simulations are representative of what would happen if the whole remnant was captured until the latest times considered.

\subsubsection{Dynamics inside the cloud}\label{sec:SN1_Internal_Dynamics}

%
%
%
%
        
            
        
%

%
%
%
%
%
\begin{figure*}
\includegraphics[width=\textwidth]{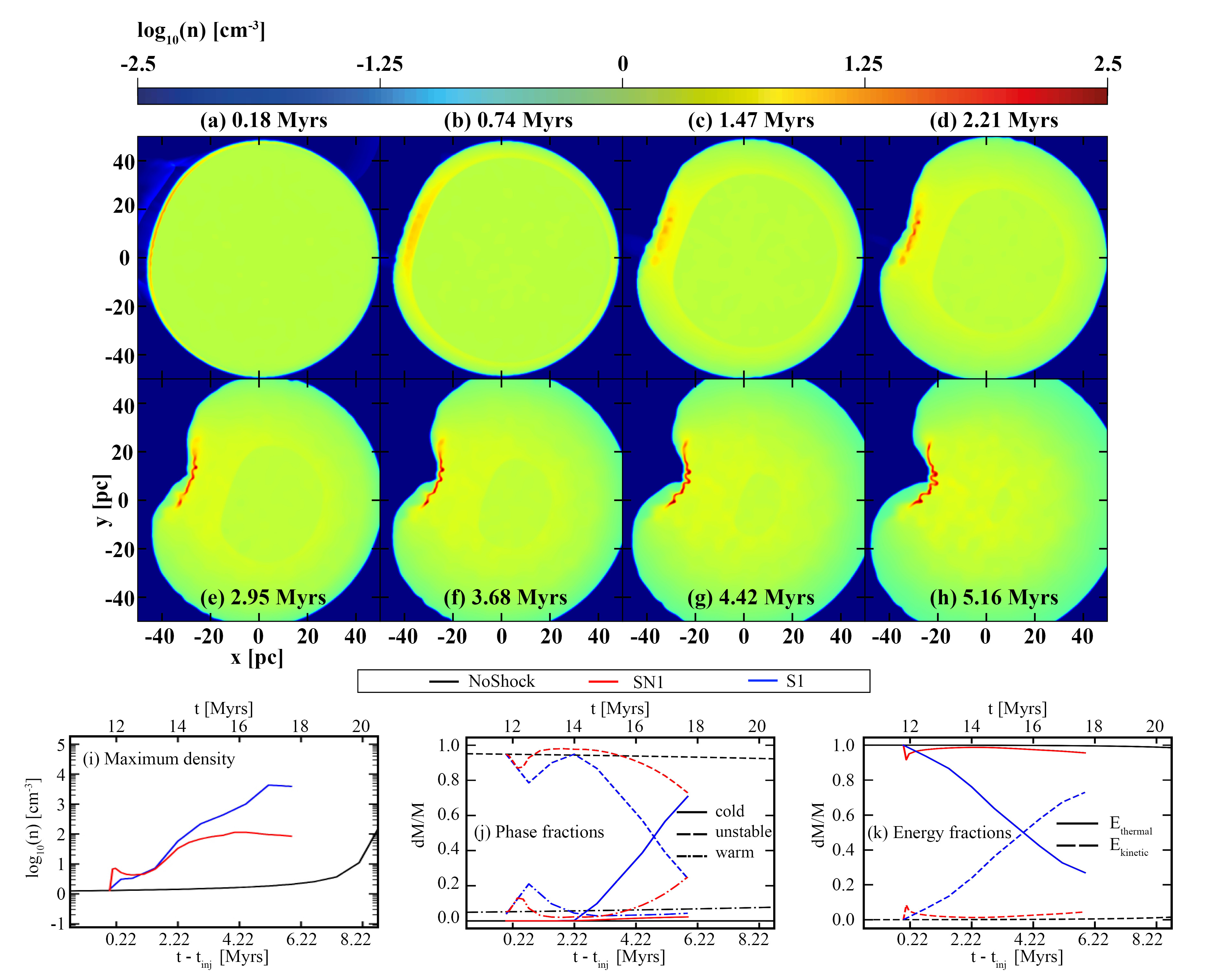}
\caption{Panels (a\,--\,h) show density slices in the \textit{z}\,=\,0 plane for the \textit{SN1} scenario. The logarithm of the number density is shown, and the time on each panel corresponds to the time elapsed since supernova injection \textit{t}\,-\,\textit{t}$_\textup{inj}$, where \textit{t}$_\textup{inj}$\,=\,11.78\,Myr. 
Panels (i), (j) and (k) respectively show the maximum density, phase fractions and energy fractions in the cloud, showing data for \textit{SN1} (red lines), \textit{S1} (blue lines) and \textit{NoShock} (black lines) for direct comparison at equivalent times. Panel (j) shows mass fractions in cold (\textit{T}\,$<$\,160\,K, solid line), unstable (160\,$<$\,\textit{T}\,$<$\,5000\,K, dashed line) and warm (5000\,$<$\,\textit{T}\,$<$\,10\,000\,K, dot-dashed line) regimes. 
Panel (k) shows the fractions of the thermal energy \textit{E}$_\textup{thermal}$ (solid line) and kinetic energy \textit{E}$_\textup{kinetic}$ (dashed line) out of \textit{E}$_\textup{tot}$\,=\,\textit{E}$_\textup{thermal}$\,+\,\textit{E}$_\textup{kinetic}$ in the cloud. Gravitational energy is ignored.}
\label{fig:SN1evo}
\end{figure*} 
%

%

%
\begin{figure*}
 \end{figure*} 
%
%
%


%
%
%
%
%
%
%
%
%
%
%
%
To describe the evolution inside the cloud, we first show snapshots of the logarithm of the density through the \textit{z}\,=\,0 plane in figure \ref{fig:SN1evo}.
Additionally, in Fig.\,\ref{fig:SN1evo}(i\,--\,k) we respectively compare the maximum density, fractions of material existing in warm (10\,000\,$>$\,\textit{T}\,$>$\,5000\,K), unstable (5000\,$>$\,\textit{T}\,$>$\,160\,K) and cold (\textit{T}\,$<$\,160\,K) thermal regimes, and the energy partition between thermal and kinetic energies.
Comparisons are drawn between the \textit{NoShock} (black line), \textit{SN1} (red line) and the \textit{S1} scenarios. 
Here, only cloud material is traced, i.e. where the passively advected scalar $\alpha$\,$>$\,0.9.
Note that in our discussions, material lying on the equilibrium curve is referred to as belonging to a phase (e.g. thermally unstable phase).
When talking about material within a temperature bracket (i.e. \textit{all} material, equilibrium + non-equilibrium), we refer to this as belonging to that particular regime (e.g. thermally unstable regime).
%
%

%
%
%
%
%
%
%
%
%
%
%
The impact of the primary shock is felt strongest on the cloud front.
%
%
As the cloud and ambient medium are initially in pressure balance, the transmitted shock has roughly equivalent strength in the cloud as it does in the surroundings, resulting in an $\textit{M}$\,$\approx$\,7 shock (\textit{M}\,=\,\textit{v}$_s$/\textit{c} where \textit{v}$_\textup{s}$\,$\approx$\,50\,km\,s$^{-1}$ and \textit{c}\,$\approx$\,7\,km\,s$^{-1}$) travelling into the cloud and thermalising a thin layer of material - as seen in Fig.\,\ref{fig:SN1evo}(a).
%
Just like for a planar shock, the cloud causes the supernova shock to diffract and reduce its strength, thus reducing the strength of the transmitted shock.
Its strength is weakened further as the remnant expansion itself slows down, which is not an effect that occurs with a planar shock due to a continuously driven flow.
Thus when the primary shock reconnects at the rear, much of its strength has been lost, with the convergence having little effect on amplifying the shock impact.
This is in stark contrast to the \textit{S1} scenario, where the strongest compression was seen at the back of the cloud due to the converging flow.
%
%
%


    
%
%
%
%
By Fig.\,\ref{fig:SN1evo}(b), the shock has travelled over the whole cloud and the transmitted shock is moving inwards from all sides.
Where the impact was the strongest, the thermalised state experiences an increase in cooling, causing the material to be compressed by the surroundings as its temperature and pressure drops.
This effect is only evident at the front of the cloud, resulting in the formation of the single cap-like structure that after 2\,--\,3 Myrs appears filament-like in the density slices shown.
The shape of this structure is distorted by the converging flow, and the pinching of the front causes the structure to accelerate faster at the centre than the sides.
This is an effect more vividly seen after $\sim$\,3\,Myrs (Fig.\,\ref{fig:SN1evo}e) and is exaggerated by the expansion of the surrounding envelope.
The remaining evolution is gradual, where the cloud continues to expand and the cold structure continues to be distorted by the converging flow.
The shock inside the cloud continues to slow down, and by Fig.\,\ref{fig:SN1evo}(h) it has still not converged at the centre, in contrast to \textit{S1} which saw the cloud shock converge on this timescale. 
%
%
%
%

%
%
%
%
%
%
Many significant differences between the \textit{S1} and \textit{SN1} scenarios can be identified and can all be explained by this fact that the continuously replenished flow in \textit{S1} provides constant ram pressure impact and constant thermal compression, and the remnant does not.
%
%
Because of this, in \textit{S1} the period between 1\,--\,4\,Myrs saw the formation of a cold thin shell spanning most of the cloud's edge which was continuously impacted by the incident flow.
%
%
This flow distorted the shell, triggering Rayleigh-Taylor and Vishniac instabilities at the front, with the sides being distorted by the Kelvin-Helmholtz instability.
Where the flow converged at the rear, the shell became winged-like, with a protruding needle forming in the centre that was subsequently accelerated against the motion of the cloud.
The surrounding flow maintained high thermal pressures, and in combination with the strong ram pressure, the cloud was continuously compressed to create conditions where gravity could take over, and collapse individual structures.
%
%
%
%
 %
%
 %
  %
%
%
%
%
%
%
Whereas here, instead of a cold thin shell forming around the edges, the only cold region is found in the small layer at the front.
%
%
%
This forms due to an instantaneous impact with no subsequent continuous flow.
As such, the ram pressure is smaller by a factor of $\sim$\,100 and thermal pressure by a factor of $\sim$\,2.5 (see Section \ref{sec:SN1external}).
None of the hydrodynamical instabilities present in \textit{S1} are triggered, and the continuous drop in pressure means that the cloud expands rather than contracts. 
The only other source of compression - the secondary shock, has no effect on the material in the cold layer and only somewhat affects the warm material at the interface. 
We do not expect to see star formation take place on the same timescales as \textit{S1}.
The stronger and more instantaneous impact of the supernova can be seen in the maximum density (Fig.\,\ref{fig:SN1evo}i), subsequently followed by a rapid decline in the next 200\,kyrs.
The phase and energy fractions (Fig.\,\ref{fig:SN1evo}j,k) also show this behaviour, with the kinetic energy specifically showing a rapid decay akin to that observed in the studies of \citet{seifried2018molecular}.
\textit{S1} overshoots \textit{SN1} in the kinetic energy after $\sim$\,300\,kyrs, maximum density after  $\sim$\,1\,Myr, and cold mass fraction after $\sim$\,2\,Myr.
The sustained compression and acceleration of material in the \textit{S1} scenario subsequently results in the steady increase of these properties, with the final maximum density growing to values of $\sim$\,10$^4$\,cm$^{-3}$, 2 orders of magnitude higher than \textit{SN1}.
At \textit{n}\,=\,10$^4$\,cm$^{-3}$, gravity is dominating, and since 70 per cent of the \textit{S1} material is in the cold regime (Fig.\,\ref{fig:SN1evo}j), the probability of star formation is high.
Indeed our results in Paper I showed evidence of local gravitational collapse.
%
At a maximum \textit{n}\,=\,100\,cm$^{-3}$ however and with less than 5 per cent of material in the cold regime, clearly in \textit{SN1} the conditions for self-gravity to dominate and cause local gravitational collapse are not created.
We note however that the cold mass fraction does appear to increase in Fig.\,\ref{fig:SN1evo}(j) (although the maximum density decreases) suggesting that the structure is not in the final stage of formation.
%
%
%
%
%

\subsubsection{Formation of cold material}


%
%
%
\begin{figure*}
\includegraphics[width=\textwidth]{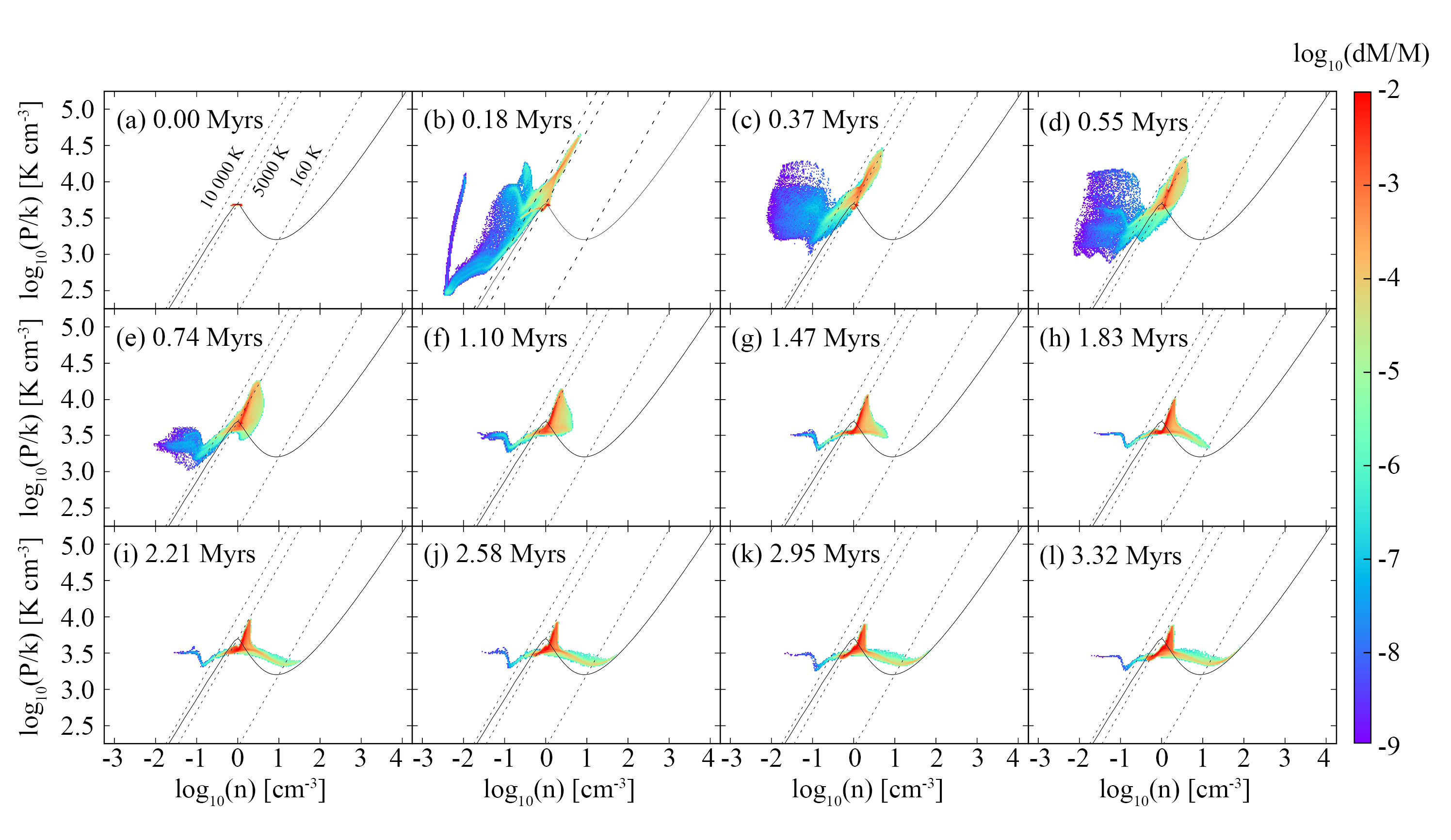}
\caption{Panels (a\,--\,l) show the mass distribution of cloud material represented in pressure-density space for the \textit{SN1} scenario. Isotherms differentiating the hot (\textit{T}\,$>$\,10\,000\,K), warm (10\,000\,$>$\,\textit{T}\,$>$\,5000\,K), unstable (5000\,$>$\,\textit{T}\,$>$\,160\,K) and cold (\textit{T}\,$<$\,160\,K) thermal regimes are shown on each panel. Presented is a period of 2.95\,Myrs from \textit{t}$_\textup{inj}$\,=\,11.78\,Myrs, after which material settles into a relatively steady state where the cold material evolves primarily due to gravity, and the warm envelope expands into the surroundings.}
\label{fig:SN1phase}
\end{figure*} 
%
%
%
%
Figure \ref{fig:SN1phase} shows the mass distribution of cloud material in pressure-density space, along with the equilibrium curve and isotherms for the hot (\textit{T}\,$>$\,10\,000\,K), warm (10\,000\,$>$\,\textit{T}\,$>$\,5000\,K), unstable (5000\,$>$\,\textit{T}\,$>$\,160\,K) and cold (\textit{T}\,$<$\,160\,K) thermal regimes. 
Note that the data shown is only for material where the advected scalar $\alpha$\,$>$\,0.9, i.e. cloud material.
%
%
%
%

%
%
%
Panel (a) shows the moment before any impact. 
Note that the remapped material exists where $\alpha$\,$<$\,0.9, and as such it is not seen in the panel.
Panel (b) clearly shows the compression of cloud gas due to the transmitted shock, and the heating of unstable material to the warm regime.
The low-pressure, low-density gas that results from the reflection of the primary shock has now mixed with the cloud and is traced by the scalar $\alpha$, and thus can be seen to exist in the hot regime.
In panel (c), a clear rise in pressure and density can be seen with this material, which is an effect that can be credited to the impact of the secondary shock.
In the warm regime, a drop in density and a broader pressure distribution can also be seen as material begins to cool towards thermal equilibrium, a pattern that continues in panels (d) and (e).
%
%
%
%
%
%
%
%
%
What is most striking is the behaviour of the gas over the next 1.5\,Myr.
In the \textit{S1} scenario, post-shock gas is seen to cool from the warm regime directly to the cold phase, completely passing the unstable phase.
Here we instead see material cool from the warm regime back to the unstable phase, and then migrate towards the cold phase from the unstable phase - in a fashion that resembles action of the thermal instability.
This is exactly what was seen in the \textit{NoShock} scenario presented in Paper I (in particular their figs. 4(a\,--\,e)).
In \textit{NoShock}, the thermal instability formed cold and dense clumps embedded in warm intercloud gas.
We would therefore expect to see such structures form here, which indeed appears to be the case.
What looks like a filament in the density slices shown in figure \ref{fig:SN1evo}(c\,--\,h) is in fact a cross-section through a broader cap-like structure, with what looks like clumps embedded in a diffuse warm envelope, features better seen in projection.
%
%
%

%
%
%
\begin{figure}
\includegraphics[width=0.5\textwidth]{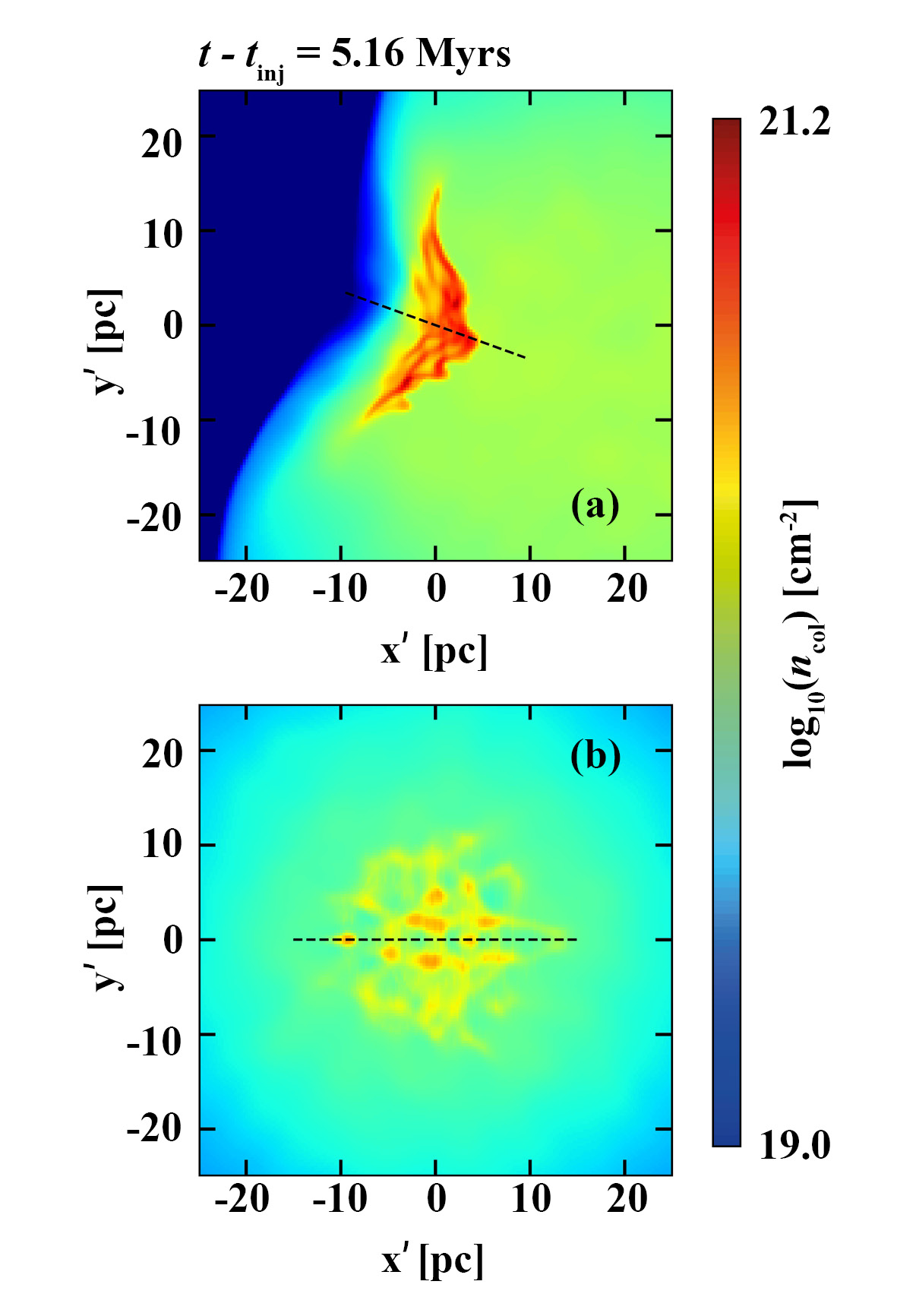}
\caption{Shown are logarithm plots of column density projections along two lines of sight for the \textit{SN1} scenario. The dashed line in panel (a) corresponds to the line of sight for panel (b), and vice versa in panel (b). The length of each line corresponds to the projection depth. Note that the axes in the image do not correspond to the original axes of the simulation, and instead form the axes of the projected image centered on the centre of mass of the cold structure (\textit{T}\,$<$\,160\,K).}
\label{fig:SN1Proj}
\end{figure} 
In Fig.\,\ref{fig:SN1Proj} we show column density projections at \textit{t}\,=\,5.16\,Myr.
Two lines-of-sight (LOS) are considered, with the image in panel (a) being a projection perpendicular to the \textit{z}\,=\,0 plane (LOS shown in panel b), and (b) being along the axis of symmetry as determined by the blast (LOS shown in panel a).
Here, the length of the lines represent the projection depth, such that only material within the volume as defined by the depth is projected.
The axes of the resultant projections are subsequently shown in the centre of mass frame of the cold material.
%
%

%
%
%
In a projection perpendicular to the \textit{x}\,--\,\textit{y} plane (Fig.\,\ref{fig:SN1Proj}a) the singular filamentary-like identity seen in Fig.\,\ref{fig:SN1evo} is mostly lost, and the projection reflects a morphology consistent with a conic, cap-like structure that has been shaped by the convergent flow.
In the projection shown in Fig.\,\ref{fig:SN1Proj}(b) we see that the filamentary identity is completely lost and within the surface there appears to be a small complex of clumps separated by distances typical of structures created by the thermal instability ($\sim$\,3-5\,pc, see WPFVL16 and Paper I). 
The clumps themselves have average densities \textit{n}\,$\approx$\,80\,cm$^{-3}$, temperatures \textit{T}\,$\approx$\,30\,--\,50\,K and sizes \textit{r}\,$\approx$\,1\,--\,2\,pc.
We thus see that the supernova shock has triggered accelerated thermal evolution in the localised area of the `front' of the cloud, closest to the supernova, and on realistic timescales formed the TI-driven structure expected from such a phase transition \citep[e.g. see ][]{falle2020thermal}.
As mentioned, the cold mass fraction is gradually increasing, suggesting that these clumps are not in their final stages of formation, however no star formation is observed.
%
%
%

\subsection{Interaction with molecular cloud - SN2}










    

        

    
    
      
      

      
      
%
%
%
%
Just like \textit{SN1}, we present the \textit{SN2} scenario by focussing first on the dynamics outside the cloud and secondly on the dynamics inside the cloud.
Throughout our discussion we compare and contrast this behaviour to the idealised planar shock scenario \textit{S2}, the un-shocked scenario \textit{NoShock}, and the previously presented scenario \textit{SN1}.
%
%
%

\subsubsection{Dynamics outside the cloud}
%
%
%
%
%
%

%

\begin{figure*}
\includegraphics[width=\textwidth]{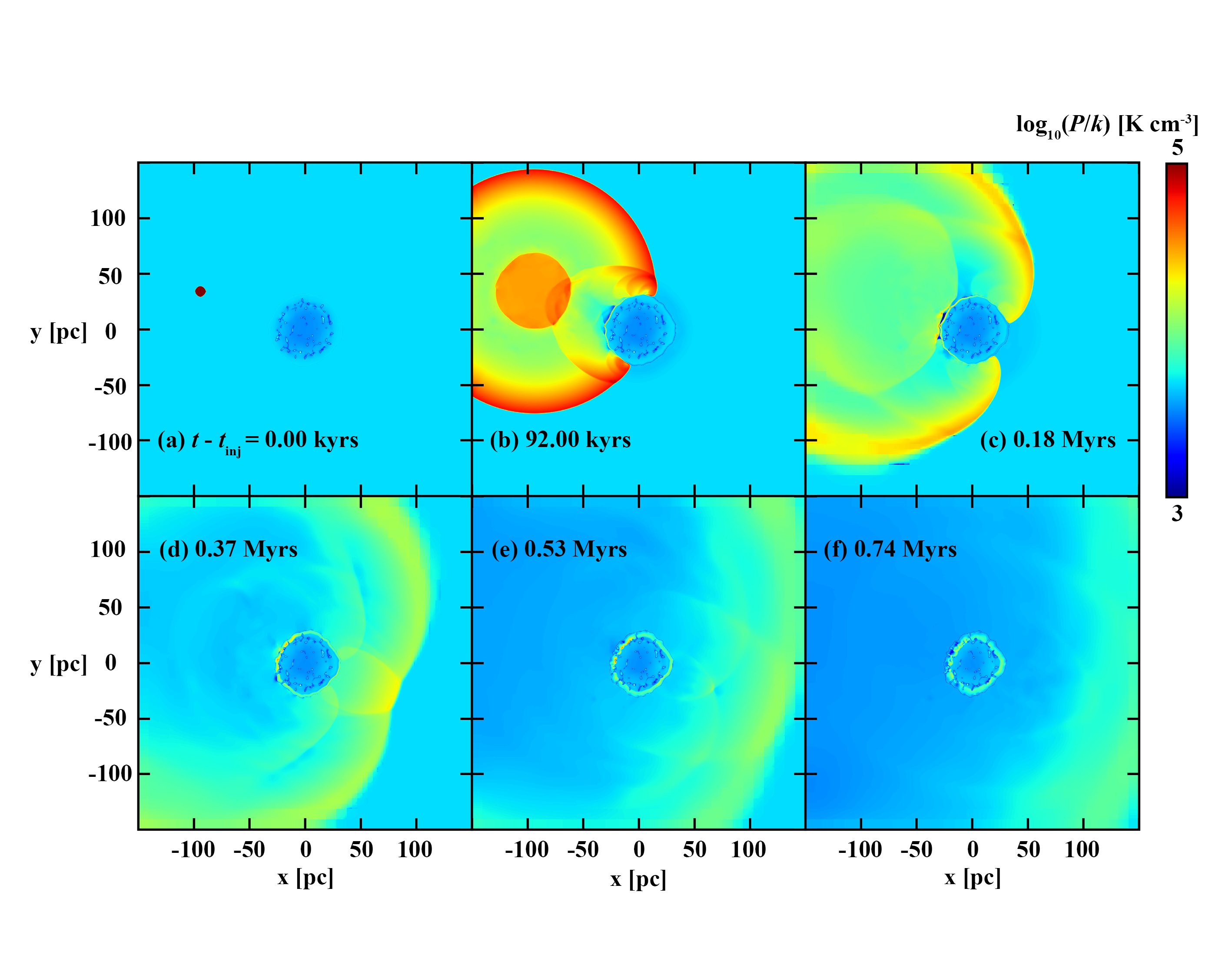}
\caption{Shown are pressure logarithm plots in the \textit{z}\,=\,0 plane of the \textit{SN2} scenario. The time in each panel corresponds to the time since supernova injection \textit{t}-\textit{t}$_\textup{inj}$, with \textit{t}$_\textup{inj}$\,=\,26.5\,Myr. Just like the \textit{SN1} scenario, the final snapshot shows the moment immediately prior to the primary shock moving off the grid and the development of the upstream flow.}
\label{fig:SN2pg}
\end{figure*}
\begin{figure*}
\includegraphics[width=\textwidth]{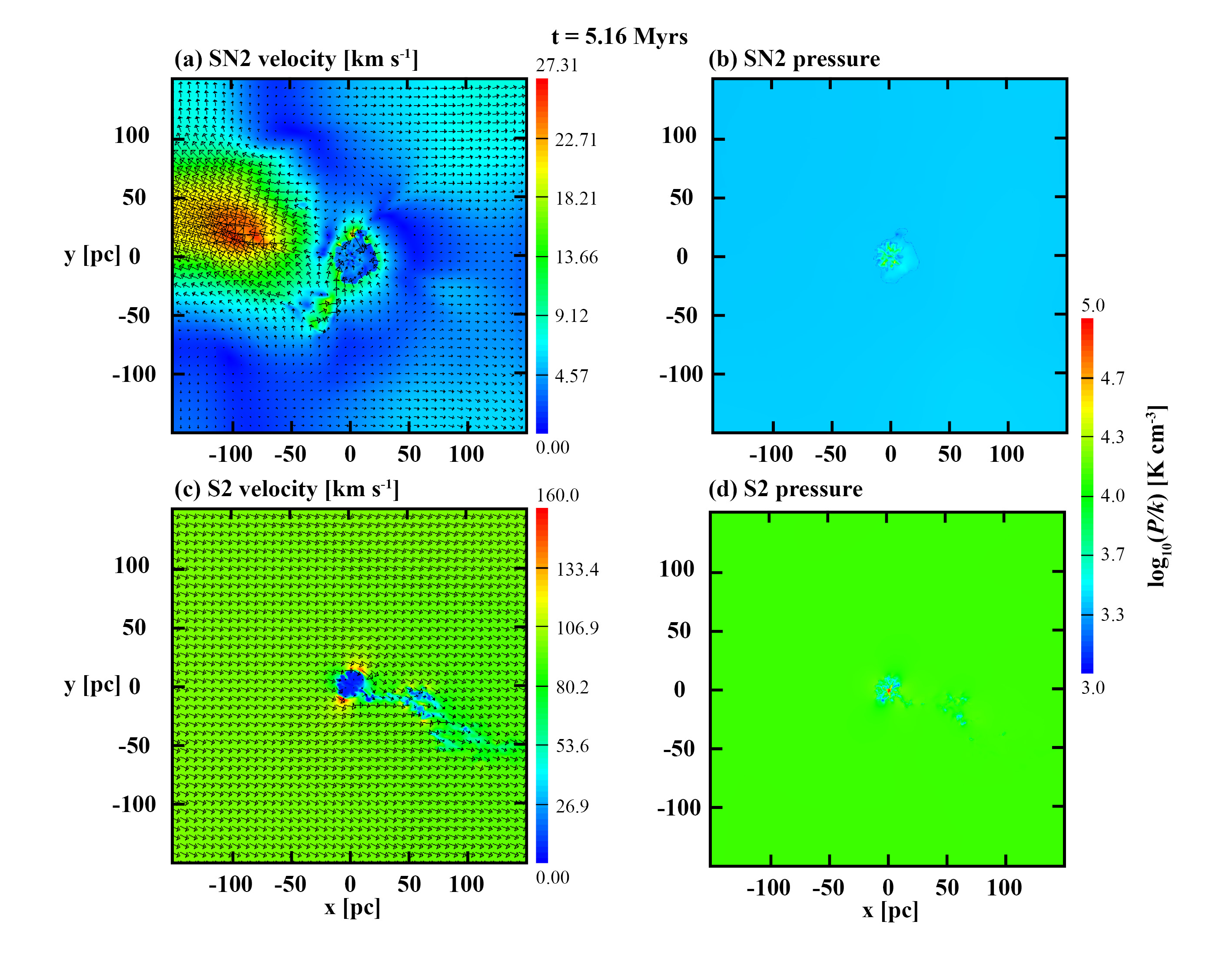}
\caption{Shown are plots in the \textit{z}\,=\,0 plane showing the velocity (panels a and c) and pressure (panels b and d) of \textit{SN2} (panels a and b) and \textit{S2} (panels c and d) at \textit{t}\,-\,\textit{t}$_\textup{inj}$\,=\,5.16\,Myrs - the final snapshot presented in this work and the timescale at which the \textit{S2} scenario witnessed local gravitational collapse.}
\label{fig:SN2velpg}
\end{figure*}

%
%
%
%
We illustrate the external behaviour by again showing snapshots of the pressure logarithm in the \textit{z}\,=\,0 plane \mbox{(figure~\ref{fig:SN2pg})} and a snapshot at 5.16\,Myrs showing pressure and velocity profiles in the same plane for both \textit{SN2} and \textit{S2} \mbox{(figure \ref{fig:SN2velpg})}.
The full evolution is presented on the same timescale of 5.16\,Myrs which just like for \textit{S1} is the timescale of the first snapshot in \textit{S2} that showed evidence of local gravitational collapse.
%
%
%

%
%
%
%
%
%
%
%
%
%
%
%
%
%
%
%
%
%
%
Fig.\,\ref{fig:SN2pg}(a) shows the moment once all of the 10\,M$_\odot$ and 10$^{51}$\,erg are injected into the domain.
A supernova explosion follows, and the remnant propagates out into the surroundings sweeping up mass and thermalising material as it does so.
Just like in \textit{SN1}, a Sedov-Taylor-like profile develops with a forward shock, reverse shock, and a contact discontinuity separating the two.
The reverse shock then reflects at the blast epicentre, sending a secondary shock propagating back towards the cloud.
%
%
%

%
%
%

            
        
%
%
%
%
%
%
%
%
The subsequent evolution is very similar to \textit{SN1}, however subtle differences are worth pointing out.
Firstly, the cloud no longer has a smooth edge, resulting in multiple wave reflections that distort both the primary and secondary shocks.
As the primary shock sweeps over the cloud and interacts with the inhomogeneities, sets of oblique shocks form which distort the profile further.
These are seen to persist for up to 180\,kyrs (Fig.\,\ref{fig:SN2pg}c), and are destroyed after the passage of the secondary shock.
%
Secondly, the cloud is smaller than in \textit{SN1}, resulting in shorter timescales for the passage of the shocks.
As can be seen in Fig.\,\ref{fig:SN2pg}(d) - the primary shock has converged, and the panel captures the moment prior to secondary shock convergence.
Contrasting this to \textit{SN1} (Fig.\,\ref{fig:SN1pg}d), the primary shock has yet to converge at this time.
In spite of these differences, in the next $\sim$\,4\,Myrs the ambient medium behaves almost identically to the \textit{SN1} scenario:
a low velocity, laminar upstream flow develops that converges at the front of the cloud, and the thermal pressure drops globally.
%
%
%

%
%
%
%
%
    
        
%
%
The final state of this behaviour is seen in figure \ref{fig:SN2velpg}, showing the same order of magnitude differences between the velocity and thermal pressures, giving a thermal pressure difference of a factor $\sim$\,2.5 lower in \textit{SN2} than \textit{S2}, and ram pressure difference a factor $\sim$\,100 lower (on the cloud edge).
The most surprising behaviour in this scenario is the state of the \textit{SN2} cloud itself, showing that although the cloud becomes immersed in under-pressured surroundings just like in \textit{SN1}, the cloud has not expanded.
The cloud in fact continues to contract (albeit not as fast as \textit{S2}), which is clearly a gravitational effect.
In this scenario therefore, the ambient behaviour is only weakly coupled to the cloud.
%
%
%
%

%
%
%
\subsubsection{Dynamics inside the cloud}

    
    

      



%

        
        
        
        
            
        
        
        
        
%
%
\begin{figure*}
\includegraphics[width=\textwidth]{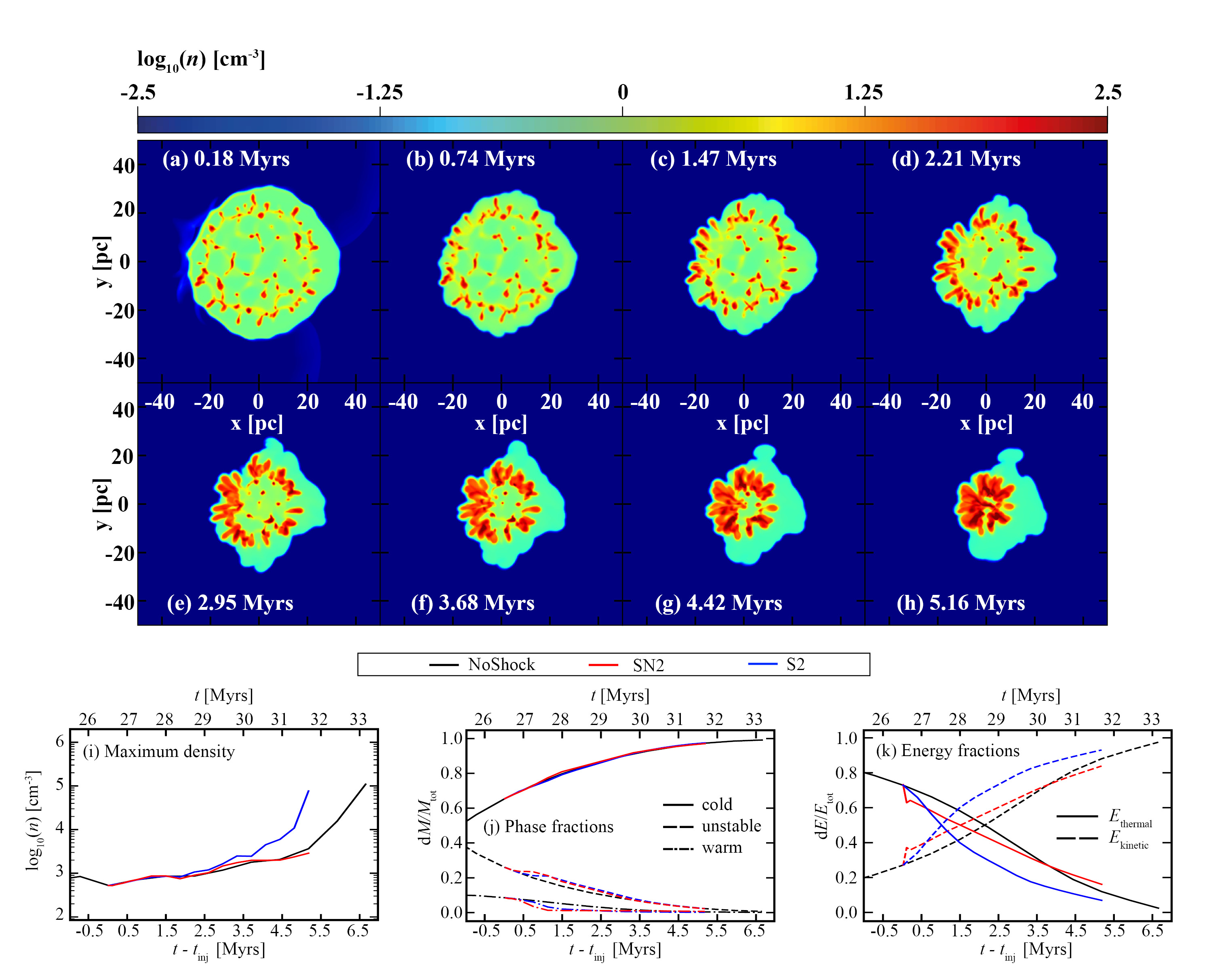}
\caption{Panels (a\,--\,h) show density slices in the \textit{z}\,=\,0 plane for the \textit{SN2} scenario. The logarithm of the number density is shown, and the time on each panel corresponds to the time elapsed since supernova injection \textit{t}\,-\,\textit{t}$_\textup{inj}$, where \textit{t}$_\textup{inj}$\,=\,26.5\,Myr. 
Panels (i), (j) and (k) respectively show the maximum density, phase fractions and energy fractions in the cloud, showing data for \textit{SN2} (red lines), \textit{S2} (blue lines) and \textit{NoShock} (black lines) for direct comparison at equivalent times. Panel (j) shows mass fractions in cold (\textit{T}\,$<$\,160\,K, solid line), unstable (160\,$<$\,\textit{T}\,$<$\,5000\,K, dashed line) and warm (5000\,$<$\,\textit{T}\,$<$\,10\,000\,K, dot-dashed line) regimes. 
Panel (k) shows the fractions of the thermal energy \textit{E}$_\textup{thermal}$ (solid line) and kinetic energy \textit{E}$_\textup{kinetic}$ (dashed line) out of \textit{E}$_\textup{tot}$\,=\,\textit{E}$_\textup{thermal}$\,+\,\textit{E}$_\textup{kinetic}$ in the cloud. Gravitational energy is ignored.}
\label{fig:SN2ev0}
\end{figure*} 
%
%
%
%
%
%
%
%
We now discuss the dynamics inside the cloud and refer to the density logarithm snapshots and statistics in figure \ref{fig:SN2ev0}. Again we show the maximum density (i), mass fractions in the cold, unstable and warm thermal regimes (j), and the fractions of thermal and kinetic energies (k).  
%
%
%
%
    

%
%
%
%
This scenario begins at 26.5\,Myr.
The evolution of the material inside the cloud has so far been dominated by the effects of the thermal instability which has generated a complex of cold and dense clumps embedded in a warm diffuse gas.
The evolution of the thermal instability in the \textit{NoShock} case was covered in section 3.1 of Paper I.
To summarise, the initial grid-scale inhomogeneities initialised around the thermally unstable equilibrium smooth out and seed the instability within the first few thousand years of evolution.
The cloud remains quiescent for $\sim$\,15\,Myr, corresponding to a period that sees a growth of the inhomogeneities, enhancing the pressure and density differences within the cloud.
A critical point is reached within the next $\sim$\,5\,Myrs where the cooling rates increase such that cold material cools further, causing it to be further compressed by the over-pressured surroundings. 
This continues to cool until temperatures reach a stable cold equilibrium, resulting in densities increasing from \textit{n}\,$\sim$\,1\,cm$^{-3}$ to \textit{n}\,$>$\,100\,cm$^{-3}$ in a period of $\sim$\,3\,Myr.
Gravity subsequently increases these densities to $\sim$\,1000\,cm$^{-3}$ prior to the global collapse of the cloud at $\sim$\,35\,Myr.
%
%

%
%
%
Following phase transition and prior to the global collapse, the state of the cloud is as a complex of cold and dense (\textit{T}\,$\sim$\,50\,--\,160\,K, \textit{n}\,$\sim$\,100\,--\,1000\,cm$^{-3}$) clumps embedded in a warm and diffuse (\textit{T}\,$\approx$\,5000\,K, \textit{n}\,$\approx$\,0.8\,cm$^{-3}$) gas.
%
%
%
%
%
Low-density, diffuse (\textit{T}\,$\approx$\,700\,K, \textit{n}\,$\approx$\,5\,cm$^{-3}$) structures resulting from thermal flows connect the clumps, and the clumps appear somewhat elongated towards one another,
serving as evidence that these clump-connecting structures are possible regions of filament formation (e.g. seen in the high resolution hydrodynamical simulations of WFP19).
As mentioned, the first detailed study of this behaviour was conducted by WPFVL16 and for extra details and comparisons between non-gravitational simulations and ones including magnetic fields - we refer the reader there.
%
%
We now describe the impact of the supernova on material inside the cloud.
%
%
%

%
    
%
%
%
%
%
%
%
%
%
Upon impact, the primary shock has an equivalent strength to that in \textit{SN1} with an approach velocity of $\sim$\,1000\,km\,s$^{-1}$.
The interclump material is impacted first and accelerated towards the cloud interior.
By Fig.\,\ref{fig:SN2ev0}(b) the shock has interacted with the clumps, and many individual shock-clump interactions ensue.
The clumps are accelerated less than their surroundings, thus interclump material is seen to be pushed inwards between the clumps - most apparent in panels (c\,--\,d).
Note that by panel (d), the upstream flow around the cloud contributes to this effect.
%
%
%
%

%
%
%
%
As the shock interacts with individual clumps, they appear to slightly elongate towards the cloud centre and develop tail-like morphologies, an effect due to their outer envelopes being stripped in the direction of the post-shock flow and potentially due to impact with the material swept up by the shock itself \citep{Pittard2011Tails}.
The material connecting the clumps is disturbed also and contributes to this tail-like appearance, again orienting radially towards the centre.
Note that the irregular interface and presence of clumps distorts the shock front, and as such its position is not obvious in the panels.
However by panel (d) it is evident the shock has interacted with outer clumps around all of the cloud's edge, as they all appear to have these elongated morphologies from their stripped envelopes.
%
%
%

%
Tail-like features were seen both in \textit{NoShock} and \textit{S2}, but to very different extents.
In the \textit{NoShock} scenario, material was accelerated radially inwards by gravity only, with clumps accelerated slower than the interclump material.
The resulting velocity gradients were quite weak, although large enough to slightly ablate the clump envelopes, resulting in similar, albeit much less pronounced tail-like morphologies compared to \textit{S2/SN2}.
This is seen in both Paper I and WPFVL16 (who also observe that equivalent models in 2D exaggerate this effect due to motion being restricted to a plane).
In the \textit{S2} scenario, this effect is much more pronounced, as after shock passage the clumps become entrained in a continuous flow. 
Although the initial impact is weaker than in \textit{SN2}, the post-shock flow is in total much more disruptive, with clump material being continuously ablated.
The constant flow results in continuous acceleration of the interclump material, resulting in some clumps breaking away from the cloud.
These clumps become fully entrained in the external flow, which further exaggerates the formation of elongated tails, and destroys some clumps completely.
In contrast, no clumps are seen to break away from the cloud in the \textit{SN2} scenario.

%
%
%
%
%
%
%
%
%
%
%

%
%
%
In addition to seeing clumps elongate, they also appear to increase in size and coalesce into larger cold regions.
This appears to be due to their outer envelopes expanding.
As the shock enters the cloud, its post shock pressure drops due to the combination of increased cooling in the cloud, gas attempting to match the decreasing external pressure, and gas expansion effects characteristic of Sedov-Taylor-like evolution.
Due to these effects, the clumps become embedded in material with lower pressure than the initial interclump material, and their outer envelopes expand, with the remainder of the material bound together by gravity.
The cooling is strongest at the front, and as such we see the majority of clump expansions occur there (see Fig.\,\ref{fig:SN2ev0}d\,--\,f).
Thus this increase in size is not to be mistaken by a gain in clump mass - also apparent by the lack of deviation from the cold mass fraction \textit{NoShock} (Fig.\,\ref{fig:SN2ev0}j).
%
%
%

%
%
%
Note that what we could also be seeing is thermal-instability induced material forming at the front in a manner similar to \textit{SN1}.
However, it is difficult to find explicit evidence that would differentiate this from the individual clump expansion, as the effects of the thermal instability would be seen at the front (similar to \textit{SN1}), which is where the clumps expand the most.
In addition, any migration across the phase diagram akin to figure \ref{fig:SN1phase} is obscured as the existence of the two-phase medium occupies exactly that portion of the diagram.
%
The only possible evidence is a short-lived increase of cold material seen in Fig.\,\ref{fig:SN2ev0}(j) which occurs on the order of the cooling timescale of the post-shock gas.
However this is negligible ($\lesssim$\,1\% increase, $\sim$\,100\,M$_\odot$), and returns almost immediately to the \textit{NoShock} fraction.
%
Thus in this scenario we conclude that once the clumps have formed, the supernova impact does not trigger any further thermal instability.
%
%
%

%
    
    
    

%
%
%
At later stages of evolution, many of the clumps appear to collide and merge.
This is interesting, as we would expect this effect to raise the maximum density in the cloud.
Inspecting individual clumps shows that densities do increase in merging regions, however these are small and do not exceed the maximum. 
There does appear to be a slight increase in the maximum density for a brief period (see Fig.\,\ref{fig:SN2ev0}i) however again this returns to \textit{NoShock} values, and eventually decreases further.
In \textit{S2}, we can see that at 5.16\,Myrs densities are attained that are two orders of magnitude larger ($\sim$\,10$^5$\,cm$^{-3}$) than in \textit{SN2} ($\sim$\,10$^3$\,cm$^{-3}$).
Once again, this can be attributed to the fact that the \textit{S2} cloud was continuously compressed, resulting in local gravitational collapse, and an acceleration of the global collapse of the cloud.
%
%
%

%
%
%
%
It is interesting therefore to observe that a supernova blast is unable to create such conditions.
%
%
In fact, looking at the density snapshots, and examining the extracted data from the model - after the first 3.5\,Myrs the cloud is almost indistinguishable from the \textit{NoShock} scenario.
%
%
%
%
%
%
Once the initial impact of the supernova is seen, the cloud settles into a steadily, globally collapsing state with gravity dominating the evolution.
Nevertheless, even with more than 50 per cent of the cloud material being cold and dense (increasing up to $\sim$\,90 per cent), a single supernova event is unable to aid gravity in collapsing individual structures prior to the global collapse of the cloud.
Thus we conclude that no extra star formation is to be expected from this interaction.
Further work is warranted to see if such behaviour is seen when a supernova interacts with TI formed clouds with different masses, and ones simulated at higher resolution that results in the hydrodynamic formation of filaments (as seen in WFP19).
\section{Discussion}\label{sec:discuss}

One would expect that all else being equal, the closer a blast source is to an object, the stronger the impact. Indeed, studying the impact of supernovae on molecular cloud velocity dispersions, \citet{seifried2018molecular} found that doubling \textit{R}$_\textup{c}$ decreased the maximum impact by roughly 40\%. It is interesting therefore to observe that shocks significantly disrupt clouds in the small cloud regime, while they do not have much impact on clouds in the large cloud regime, given that the small cloud regime is often employed as an approximation to supernova remnants. It is prudent to ask how reflective such simulations are of realistic astrophysical systems.

To illustrate such considerations with our clouds, a density contrast $\chi$\,$\approx$\,500 evaluates equation (4) to \textit{r}$_\textup{cl}$\,$\ll$\,0.0047\textit{R}$_\textup{c}$, requiring a cloud to have a radius that is approximately 0.5\,\% the size of the blast radius \textit{R}$_\textup{c}$ to qualify for the small cloud regime. For a cloud with \textit{r}$_\textup{cl}$\,=\,50\,pc this requires a minimum \textit{R}$_\textup{c}$\,$\approx$\,11\,kpc, which is unrealistic for a single supernova. In fact, using equation (39.31) in \citet{draine2010physics} to calculate the "fade-away radius" - the theoretical length-scale at which the shock Mach number falls below unity and a supernova remnant mixes with the ISM, we get \textit{R}$_\textup{fade}$\,$\approx$\,200\,pc for our 10\,M$_\odot$, 10$^{51}$\,erg supernova exploding into a medium with \textit{n}$_\textup{amb}$\,=\,0.0022\,cm$^{-3}$. Using \textit{R}$_\textup{fade}$ to approximate the maximum distance \textit{R}$_\textup{c}$ that would lead to an interaction, according to equation (4), only clouds with \textit{r}$_\textup{cl}$\,$\ll$\,1\,pc would fall into the small cloud regime. Clearly, our \textit{S1/S2} models are not representative of a supernova like that seen in \textit{SN1/SN2} exploding far away.

Whilst a single supernova may not drive the kind of wind seen in \textit{S1}/\textit{S2}, a superbubble due to a cluster of supernovae might, opening up the parameter space to exploring regimes between a single supernova, to whole clusters of supernovae \citep[e.g.][]{padoan2016supernova,fielding2018clustered}. For medium and large cloud regimes, it is clear that a constraint of the parameter space requires a supernova rate of 1 per \textit{t}$_\textup{p}$ (equation 1) as a necessary condition in a clustered environment. Given the size of our clouds, it is likely that our \textit{S1}/\textit{S2} scenarios are thus more reflective of the kinds of clouds embedded in galactic winds and superbubbles. 

To speculate further however, given typical wind speeds (500\,--\,1500\,km\,s$^{-1}$) and  molecular phase speeds (50\,--\,300\,km\,s$^{-1}$) \citep[see e.g.][]{strickland2009supernova,rupke2018review} it's unclear at what region of the parameter space these models occupy. 
For instance, at 5.16\,Myrs when evidence of local gravitational collapse was seen and the simulations were stopped, the cloud's centre of mass velocities were only $\sim$\,6\,km\,s$^{-1}$. Therefore it does not appear likely that our clouds could be accelerated to speeds close to 50\,km\,s$^{-1}$ before the cloud is destroyed by the external wind or subsequent feedback.

With this in mind, it is nevertheless possible that in a scenario with feedback, a powerful wind due to a star with mass $>$ 100 M$_\odot$ breaks out \citep[e.g.][]{rogers2013feedback,wareing2017hydrodynamic} and mass loads the external flow. In such a scenario, warm cloud material could be expelled out of the cloud and subsequently cool and condense, perhaps due to the thermal instability, forming fast moving clumps and resulting in a system with a hot ionized component and a cold molecular component \citep[see e.g.][]{zhang2017entrainment,gronke2018growth}. However at this stage this behaviour is very speculative, and further exploration of the parameter space is needed to draw any more conclusions. This is beyond the scope of this paper, and is left for future work.

\section{Conclusion}\label{sec:conclucions}
In this work, we study the interaction of a supernova remnant with a cloud that is evolving due to the thermal instability (TI) and gravity, and draw comparisons to idealised shock-cloud scenarios and a scenario without shocks.
A total of five 3D hydrodynamical simulations were performed: an un-shocked scenario \textit{NoShock}, two shock-cloud scenarios in the small cloud regime, \textit{S1/S2}, respectively analogous to \textit{12Shock/24Shock} in Paper I, and two supernova-cloud scenarios in the large cloud regime, \textit{SN1/SN2}, looking at impacts at equivalent times to \textit{S1/S2}.
Just like Paper I, the disturbance is introduced at a "pre-TI" (\textit{S1/SN1}) and "post-TI" (\textit{S2/SN2}) stage, and the models are evolved for a timescale corresponding to the first snapshots when the shock-cloud scenarios \textit{S1}/\textit{S2} showed evidence of local gravitational collapse. 
For our chosen parameters, the Mach number of the supernova shock is roughly a factor $\sim$\,4.5\,$\times$ stronger at impact than that of the planar shock. However on aggregate the supernova is significantly less disruptive than the sustained impact from the idealised post-shock flow. 
In \textit{S1/S2}, the post-shock flow is constantly replenished and approaches the cloud at $\sim$\,100\,km\,s$^{-1}$. This provides a ram pressure that consistently throughout the simulation is a factor $\sim$\,100\,$\times$ larger than from the $\sim$\,10\,km\,s$^{-1}$ flows that develop around the cloud in the \textit{SN1/SN2} scenarios.
Additionally, the constantly replenished flow means that the ambient thermal pressure is maintained at higher values than those in the cloud, resulting in continuous compression throughout the entire simulation. 
In contrast, the profile within the remnant is Sedov-Taylor-like. 
As such the impact of the blast on the cloud decays rapidly, the cloud is not met with a continuous flow, and the ambient thermal pressure drops below that of the cloud in less than 1\,Myr.
%
These facts accounts for all of the differences seen between the supernova/shock models, and we observe the following particular outcomes:
\begin{enumerate}
    \item \textit{SN1}/\textit{SN2}: The cloud is impacted by two shocks - a primary shock that is the forward shock from the initial blast, and a secondary shock resulting from the reverse shock that reflects at the blast epicentre and propagates back outwards and towards the cloud. 
    The impact of the secondary shock on each cloud is weak, and its influence is mainly felt at the cloud\,--\,ambient interface.
    \item \textit{SN1/SN2}: The primary and secondary shocks sweep over the cloud and generate complex wave patterns that propagate both downstream and upstream. The upstream components generate a flow on the cloud interface that converges at the front. 
    \item \textit{SN1/SN2}: The primary shock loses much of its strength as it sweeps over the cloud.
    When it converges at the rear of the cloud, the convergence has negligible effect in amplifying the impact. In fact, as the model evolves, the strongest ram pressures occur at the front of the cloud where the upstream interface flow converges. This is in stark contrast to \textit{S1/S2}, where the strongest compression was at the rear.
    \item \textit{SN1}: The lower ambient pressure compared to that of the cloud (ram + thermal) results in a constant and gradual cloud expansion. Where the upstream flow converges, the expansion is prevented - giving the cloud a lobe-like morphology.
    \item \textit{SN1}: The lack of continuous flow means that the Rayleigh-Taylor, Vishniac and Kelvin-Helmholtz instabilities are not triggered.
    \item \textit{SN1}: The TI is triggered behind the primary shock. This forms a cap-like structure with clumps embedded inside with properties expected from such a phase transition. The maximum densities in these structures do not exceed 100\,cm$^{-3}$, however the cold mass fraction is increasing, suggesting the structure is not in its final stages of formation. Nevertheless, we do not expect star formation to occur on the same timescales as \textit{S1}.
    \item \textit{SN2}: The passage of the shock through the interclump medium results in the formation of tail-like structures due to clump envelopes being ablated. This does not happen to the same extent as in \textit{S2}, e.g. no clumps are seen to break away from the parent cloud and get entrenched in an external flow. This effect is however more pronounced than that due to flows resulting only from gravitational acceleration of material in the \textit{NoShock} scenario.
    \item \textit{SN2}: After shock passage, the clumps appear to increase in size. This is not due to an increase in mass, and instead due to expansion. This happens because the inter-clump pressure drops below that of the clumps after shock passage. 
    \item \textit{SN2}: Clumps are seen to coalesce and merge. This however does not increase the maximum density in the cloud or cause any local gravitational collapse.
    \item \textit{SN2}: Although the exterior pressure drops below that of the cloud within the first 1\,Myr, the cloud does not expand as it is gravitationally dominated. After $\sim$\,3.5\,Myr, the cloud is almost indistinguishable from the \textit{NoShock} scenario. Thus, apart from the trivial global collapse of the cloud, we do not expect to see any additional star formation.
    
\end{enumerate}



\section*{ACKNOWLEDGEMENTS}

We thank the anonymous referee for her/his helpful comments that led to the further development of the discussion section, and the physical implications of the models.
MMK acknowledges support from the Science and Technology
Facilities Council (STFC) for PhD funding (STFC DTP grant
ST/R504889/1). JMP and SAEGF acknowledge STFC for project
funding (STFC Research Grant ST/P00041X/1). CJW thanks Leeds Institute for Fluid Dynamics and the School of Chemical and Process Engineering for the visiting research fellowship. The calculations
herein were performed on ARC3, part of the High Performance
Computing facilities at the University of Leeds, UK. Additional simulations were performed using the DiRAC Data Intensive service at
Leicester, operated by the University of Leicester IT Services, which
forms part of the STFC DiRAC HPC Facility (www.dirac.ac.uk).
The equipment was funded by BEIS capital funding via STFC capital grants ST/K000373/1 and ST/R002363/1 and STFC DiRAC
Operations grant ST/R001014/1. DiRAC is part of the National e-Infrastructure. MMK would like to thank John Holden for fruitful discussions.

\section*{Data Availability}

The dataset associated with this article will be made available in the Research Data Leeds Repository at https://doi.org/10.5518/1130



\bibliographystyle{mnras}
\bibliography{bibliography} 
\end{document}